\documentclass{amsart} 
\usepackage{graphicx}
\usepackage{amssymb, amsmath, sidecap}
\usepackage{amsfonts}
\usepackage{amssymb}
\usepackage{float}
\usepackage[version=3]{mhchem}
\usepackage{endnotes}

\newcommand{\R}{\mathbb R}

\def\la{\label}

\def\R{\mathbb R}

\def\ei{\end{itemize}}
\def\bi{\begin{itemize}}
\def\bt{\begin{thm}}
\def\et{\end{thm}}
\def\bl{\begin{lem}}
\def\el{\end{lem}}
\def\bd{\begin{defi}}
\def\ed{\end{defi}}
\def\bc{\begin{cor}}
\def\ec{\end{cor}}
\def\bp{\begin{proof}}
\def\ep{\end{proof}}
\def\br{\begin{rem}}
\def\er{\end{rem}}

\def\bpp{\begin{proof}}
\def\epp{\end{proof}}

\def\bcon{\begin{conclusion}}
\def\econ{\end{conclusion}}

\newtheorem{thm}{Theorem}[section]
\newtheorem{lem}{Lemma}[section]
\newtheorem{defi}{Definition}[section]

\newtheorem{rem}{Remark}[section]
\newtheorem{cor}{Corollary}[section]

\newtheorem{conclusion}{Physical Conclusion}[section]

\numberwithin{equation}{section}
\numberwithin{figure}{section}

\begin{document}
\title{Dynamic Transition and Pattern Formation for Chemotactic Systems}
\author[Ma]{Tian Ma}
\address[TM]{Department of Mathematics, Sichuan University,
Chengdu, P. R. China}

\author[Wang]{Shouhong Wang}
\address[SW]{Department of Mathematics,
Indiana University, Bloomington, IN 47405}
\email{showang@indiana.edu, http://www.indiana.edu/~fluid}

\thanks{The authors are grateful to B. Perthame   for insightful discussions  and suggestions. The work was supported in part by grants from the
Office of Naval Research, the US National Science Foundation, and the Chinese National Science Foundation.}


\begin{abstract}
The main objective of this article is to study the dynamic transition and pattern formation for chemotactic systems modeled by the Keller-Segel equations. We study chemotactic systems with either  rich or moderated stimulant supplies. For the rich stimulant chemotactic system, we show that the chemotactic system always undergoes a Type-I or Type-II dynamic transition from the homogeneous state to steady state solutions. The  type of transition is dictated by the  sign of a non dimensional parameter $b$. For the general  Keller-Segel model where the stimulant is moderately supplied, the system can undergo a dynamic transition to either steady state patterns or spatiotemporal oscillations. From the pattern formation point of view, the formation and the mechanism of  both the lamella and rectangular patterns  are derived. 
\end{abstract}
\maketitle

\section{Introduction}
Chemotaxis is a remarkable phenomenon occurring in many biological
individuals, which involves mobility and aggregation of the species
in two aspects: one is random walking, and the other is the
chemically directed movement. For example, in the slime mould
Dictyostelium discoideum, the single-cell amoebae move towards
regions of relatively high concentration of a chemical called
cylic-AAMP which is secreted by the amoebae themselves. Many
experiments demonstrate that under some properly conditions a
bacterial coloy can form a rather regular pattern, which is relatively
stable in certain time  scale. A series of  experimental
results on the patterns formed by the bacteria Escherichia
coli (E. coli) and Salmonella typhimurium (S. Typhimurium) were
derived in \cite{BB91,BB95}, where 
two types of experiments were conducted: one is in semi-solid medium, and
the other is in liquid medium. They showed that when the bacteria
are exposed to intermediates of TCA cycle, they can form various
regular patterns, typically as ringlike and sunflowerlike
formations. In all these experiments, the bacteria are known to
secrete aspartate, a potent chemoattractant; also see \cite{murray, brenner}.

The most interesting work done by Budrene and Berg are the semi-solid
experiments with E. coli and S. typhimurium.  A high
density bacteria were inoculated in a petri dish containing a
uniform distribution of stimulant in the semi-solid medium, i.e.
0.24$\%$ soft agar in succinate. The stimulant provides main food
source for the bacteria. In a few days, the bacteria spread out from
the inoculum, eventually covering the entire surface of the dish
with a stationary pattern where the higher density population is
separated by regions of near zero cell density. The S. typhimurium
patterns are concentric rings and are either continuous or spotted;
see Figure~\ref{f11.1}. The E. coli patterns are more complex with symmetry between individual aggregates. A large
number of patterns has been observed. The most typical forms are
concentric rings, sunflower type spirals, radial stripes, radial
spots and chevrons. In the process of pattern formation, the
population of bacteria has gone through many generations.
\begin{figure}
  \centering
  \includegraphics[width=0.4\textwidth]{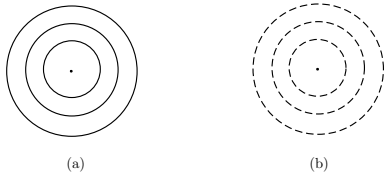}
  \caption{The black ring line and spot represent high density of
bacteria.}\la{f11.1}
 \end{figure}

The liquid experiments with E. coli and S. typhimurium exhibit
relatively simple patterns which appear quickly in a few minutes,
and last about half an hour before disappearing. Two types of
patterns are observed, and they rely on the initial conditions. The
simplest patterns are produced when the liquid medium contains a
uniform distribution of bacteria and a small amount of the TCA cycle
intermediate. The bacteria collect in aggregates of about the same
size over the entire surface of the liquid. The second type of
patterns appears when a small amount of TCA is added locally to a
special spot in a uniform distribution of bacteria. In this case,
the bacteria begin to form aggregates which occur on a ring centered
about the special spot, and in a random arrangement inside the ring.
In particular, in these liquid experiments, the timescale to form
patterns is less than the time required for bacterial birth and
death. Therefore, the growth of bacteria does not contribute to the
pattern formation process.

Here  we have to address that in these experiments, none of the
chemicals placed in the petri dish is a chemo-attractant. Hence, the
chemoattractants, which play a crucial role in bacterial chemotaxis,
are produced and secreted by the bacteria themselves.

In their pioneering work \cite{KS70},  E. F. Keller and L. A. Segel proposed a model
in 1970, called the Keller-Segel equations, to describe the
chemotactic behaviors of the slime mould amoebae. 
In their equations, the growth rate of amoeba cells
was ignored, i.e., the model can only  depict the chemotaxis process
in a small timescale, as exhibited in the liquid medium experiments
with E. Coli and S. Typhimurium by \cite{BB91,BB95}.
However, in the semi-solid medium experiments, the timescale of
pattern formation process is long enough to accommodate many
generations of bacteria. Therefore, various revised models  were
presented by many authors, taking into consideration the effects of the
stimulant (i.e. food source) and the growth rate of population; see among others \cite{murray} and the
references therein. Also, 
there is a vast literature on the mathematical studies for the Keller-Segel model; see among others \cite{perthame11, guo10, perthame09, perthame08}.

The main objective of this article is to study the dynamic transition and pattern formation for chemotactic systems modeled by the Keller-Segel equations. The  study is based on the dynamic transition theory developed recently by the authors. The key philosophy for the dynamic transition theory is to search for all transition states. The stability and the basin of attraction of the transition states provide naturally  the mechanism of pattern formation associated with chemotactic systems. 

Another important ingredient of the dynamic transition theory is the introduction of a dynamic classification scheme of transitions, with which phase transitions are classified into three types: Type-I, Type-II and Type-III. In more mathematically intuitive terms, they are called continuous, jump and mixed transitions respectively. Basically, as the control parameter passes the critical threshold, the transition states stay in a close neighborhood of the basic state for a Type-I transition, are outside of a neighborhood of the basic state for a Type-II (jump) transition. For the Type-III transition, a neighborhood is divided into two open regions with a Type-I transition in one region, and a Type-II transition in the other region. 
 

Two types of Keller-Segel models are addressed in this article. The first is the model for rich stimulant chemotactic systems (with rich nutrient supplies). In this case, the equations are a two-component system, describing the evolution of the population density of biological individuals  and  the chemoattractant concentration. In this case we show that the chemotactic system always undergoes a Type-I or Type-II dynamic transition from the homogeneous state to steady state solutions. The  type of transition is dictated by the  sign of a nondimensional parameter $b$. For example, in a a non-growth system in a narrow domain, for the spatial scale smaller than a critical number, the system undergoes a Type-I (continuous) transition,  otherwise the system undergoes a Type-II (jump) transition, leading to a more complex pattern away from the basic homogeneous state.   

The second is a more general Keller-Segel model where the stimulant is moderately supplied. In this case, the model is a three-component system describing the evolution of the population density of biological individuals, the  
chemoattractant concentration, and the stimulant concentration. In this case, the system can undergo a dynamic transition to either steady state patterns or spatiotemporal oscillation. In both cases, the transition can be either a Type-I or Type-II dictated respectively by two  nondimensional parameter $b_0$  and $b_1$.

For simplicity, we consider in this article only the case where the first eigenvalue of  the linearized problem around the homogeneous pattern is simple (real or complex), and we shall explore more general case elsewhere. In the case considered, for the Type-I transition, when  the linearized eigenvalue is simple, we show that both the lamella and rectangular pattern  can form depending on the geometry of the spatial domain. Namely,  for narrow domains, the lamella pattern forms, otherwise the rectangular pattern occurs. For Of course, for Type-II transitions, more complex patterns emerge far from the basic homogeneous state. 

The paper is arranged as follows. Section 2 introduces the Keller segel model. The rich stimulant case is addressed in Section 3, 
and the general there-component system is studied in Section 4. Section  5 explores some biological conclusions of the main theorems. 

\section{Keller-Segel Model}
The general form of the revised Keller-Segel model is given by 
\begin{equation}
\begin{aligned}
&\frac{\partial u_1}{\partial t}=k_1\Delta u_1-\chi\nabla (u_1\nabla
u_2)+\alpha_1u_1\left(\frac{\alpha_2u_3}{\alpha_0+u_3}-u^2_1\right),\\
&\frac{\partial u_2}{\partial t}=k_2\Delta
u_2+r_1u_1-r_2u_2,\\
&\frac{\partial u_3}{\partial t}=k_3\Delta u_3-r_3u_1u_3+q(x), 
\end{aligned}
\label{(11.3.13)}
\end{equation}
where  $u_1$ is the population density of  biological
individuals, $u_2$  is the chemoattractant concentration, $u_3$  is the
stimulant concentration, $q(x)$ is the nutrient source, and $\chi$  is a chemotactic response coefficient.

Equations (\ref{(11.3.13)}) are  supplemented with  the Neumann condition:
\begin{equation}
\frac{\partial (u_1, u_2, u_3)}{\partial n}=0\ \text{on}\ \partial\Omega.\label{(11.3.14)}
\end{equation}

For simplicity, we consider in this article the case where the spatial 
domain $\Omega$ is a two-dimensional (2D)  rectangle:
$$\Omega =(0,l_1)\times (0,l_2)\ \ \ \ \text{for}\ l_1\neq l_2.$$
It is convenient to introduce the nondimensional form of the model. For this purpose, let 
\begin{equation}
\begin{aligned}
&t=t^{\prime}/r_2,\quad x=\sqrt{k_2/r_2}x^{\prime},\\
& u_1=\sqrt{\alpha_2}u^{\prime}_1,\quad u_2=k_2u^{\prime}_2/\chi ,\quad  u_3=\alpha_0u^{\prime}_3,
\end{aligned}
\label{(11.3.19)}
\end{equation}
and we define the following non dimensional parameters:
\begin{equation}
\begin{aligned}
&\lambda =r_1\sqrt{\alpha_2}\chi /r_2k_2,\ \ \ \ \alpha
=\alpha_1\alpha_2/r_2,\ \ \ \ \ \mu =k_1/k_2,\\
&r=k_3/k_2,\ \ \ \ \ \ \ \ \ \ \ \ \ \ \delta
=r_3\sqrt{\alpha_2}/r_2,\ \ \ \ \delta_0=q/r_2\alpha_0.
\end{aligned}
\label{(11.3.20)}
\end{equation}
Then suppressing the primes, the non-dimensional form of  the Keller-Segel model is given by:
\begin{equation}
\begin{aligned}
&\frac{\partial u_1}{\partial t}=\mu\Delta u_1-\nabla (u_1\nabla
u_2)+\alpha u_1\left(\frac{u_3}{1+u_3}-u^2_1\right),\\
&\frac{\partial u_2}{\partial t}=\Delta u_2-u_2+\lambda u_1,\\
&\frac{\partial u_3}{\partial t}=r\Delta u_3-\delta
u_1u_3+\delta_0,\\
&\left.\frac{\partial u}{\partial n}\right|_{\partial\Omega}=0,\\
&u(0)=u_0\ \text{in}\ \Omega .
\end{aligned}
\label{(11.3.21)}
\end{equation}

The non-dimensional of $\Omega$ is written as
$$\Omega =(0,L_1)\times (0,L_2)\ \ \ \ \text{with}\ L_1\neq L_2.$$

Often times, the following form of the Keller-Segel equations is
discussed in some literatures:
\begin{equation}
\begin{aligned}
&\frac{\partial u_1}{\partial t}=\mu\Delta u_1-\nabla (u_1\nabla
u_2)+\alpha u_1\left(\frac{u_3}{1+u_3}-u^2_1\right),\\
&\frac{\partial u_3}{\partial t}=r\Delta u_3-\delta
u_1u_3+\delta_0,\\
&-\Delta u_2+u_2=\lambda u_1.\\
&\left.\frac{\partial u}{\partial n}\right|_{\partial\Omega}=0,\\
&u(0)=u_0. \end{aligned} \label{(11.3.22)}
\end{equation}

The biological significance of (\ref{(11.3.22)}) is that the
diffusion and degradation of the chemoattractant secreted by the
bacteria themselves are almost balanced by their production.  The main 
advantage of (\ref{(11.3.22)}) lies in its mathematical simplicity,  and as we shall see from the main results of this article,  the  the main characteristics of the pattern  formation associated with the model  are
retained.

\section{Dynamic Transitions for Rich Stimulant System}
\subsection{The  model}
We know that as nutrient $u_3$ is richly supplied, the
Keller-Segel model  (\ref{(11.3.13)}) is reduced to a
two-component system:
\begin{equation}
\begin{aligned}
&\frac{\partial u_1}{\partial t}=\mu\Delta u_1-\nabla (u_1\nabla
u_2)+\alpha u_1(1-u^2_1),\\
&\frac{\partial u_2}{\partial t}=\Delta u_2-u_2+\lambda u_1,\\
&\left.\frac{\partial (u_1, u_2)}{\partial n}\right|_{\partial\Omega}=0,\\
&u(0)=u_0.
\end{aligned}
\label{(11.3.29)}
\end{equation}
It is easy to see that  $u^*=(1,\lambda )$ is a steady state of (\ref{(11.3.29)}). Consider the deviation from
 $u^*$:
 $$u=u^\ast + u'. $$
Suppressing the primes,  the system (\ref{(11.3.29)}) is then transformed into
\begin{equation}
\begin{aligned}
&\frac{\partial u_1}{\partial t}=\mu\Delta u_1-2\alpha u_1-\Delta
u_2-\nabla (u_1\nabla u_2)-3\alpha u^2_1-\alpha u^3_1,\\
&\frac{\partial u_2}{\partial t}=\Delta u_2-u_2+\lambda u_1,\\
&\left.\frac{\partial (u_1, u_2)}{\partial n}\right|_{\partial\Omega}=0,\\
&u(0)=u_0.
\end{aligned}
\label{(11.3.30)}
\end{equation}

\subsection{Dynamic transition and pattern formation for the diffusion and degradation balanced case}
We start with an  important case  where the diffusion and degradation of the chemoattractant secreted by the bacteria themselves are almost balanced by their production. In this case, the second equation in (\ref{(11.3.30)}) is given by 
$$0=\triangle u_2-u_2 + \lambda u_1.$$
With the Newman boundary condition for $u_2$, we have $u_2=[-\triangle +1]^{-1} u_1$  and 
the functional form of the resulting equations are given by 
\begin{equation}
\frac{\partial u_1}{\partial
t}={\mathcal{L}}_{\lambda}u_1+G(u_1,\lambda ),\label{(11.3.33)}
\end{equation}
where the operators $\mathcal L_\lambda:H_1 \to H$ and $G: H_1 \times \mathbb R \to \R$  are defined by 
\begin{equation}
\begin{aligned}
&{\mathcal{L}}_{\lambda}u_1=\mu\Delta u_1-2\alpha u_1-\lambda\Delta
[-\Delta +I]^{-1}u_1,\\
&G(u_1,\lambda )=-\lambda \nabla (u_1\nabla [-\Delta
+I]^{-1}u_1)-3\alpha u^2_1-\alpha u^3_1.
\end{aligned}
\label{(11.3.34)}
\end{equation}
Here the two Hilbert spaces $H$  and $H_1$  are defined by 
$$
H=L^2(\Omega), \qquad H_1 = \{ u_1 \in H^2(\Omega) \ | \ \frac{\partial u_1}{\partial n}=0 \text{ on } \Omega\}.
$$

To study the dynamic transition of this  problem, we need to consider the linearized eigenvalue problem
 of (\ref{(11.3.33)}):
\begin{equation}
{\mathcal{L}}_{\lambda}e=\beta (\lambda )e.\label{(11.3.35)}
\end{equation}
 Let $\rho_k$ and $e_k$ be the eigenvalues and
eigenfunctions of $-\Delta$ with the Neumann boundary condition given by
\begin{equation}
e_k=\cos\frac{k_1\pi x_1}{L_1}\cos\frac{k_2\pi x_2}{L_2}, \qquad 
\rho_k=\pi^2\left(\frac{k^2_1}{L^2_1}+\frac{k^2_2}{L^2_2}\right), 
\label{(11.3.37)}
\end{equation}
for any  $k=(k_1,k_2)\in\mathbb N_+^2$. Here $\mathbb N_+$ is the set of 
all nonnegative integers. In particular, $e_0=1$ and $\rho_0=0$.

Obviously, the functions in (\ref{(11.3.37)}) are also 
eigenvectors of (\ref{(11.3.35)}), and the corresponding eigenvalues
$\beta_k$ are 
\begin{equation}
\beta_k(\lambda )=-\mu\rho_k-2\alpha
+\frac{\lambda\rho_k}{1+\rho_k}.\label{(11.3.39)}
\end{equation}
Define a critical parameter by
\begin{equation}
\lambda_c=\min_{\rho_k}\frac{(\rho_k+1)(\mu\rho_k+2\alpha
)}{\rho_k}.\label{(11.3.40)}
\end{equation}
Let 
$$\mathcal S=\left\{ K=(K_1, K_2) \in \mathbb N_+^2 \text{ achieves the minimization in (\ref{(11.3.40)})}\right\}.
$$
Then it follows from (\ref{(11.3.39)}) and (\ref{(11.3.40)}) that
\begin{align}
&
\beta_K(\lambda )\left\{\begin{array}{ll} <0 &\text{ if } \lambda <\lambda_c\\
=0& \text{ if } \lambda =\lambda_c\\
>0& \text{ if } \lambda >\lambda_c
\end{array}\right.   
&&   \forall K=(K_1, K_2) \in \mathcal S,\label{(11.3.41)}
\\
&
\beta_k(\lambda_c)<0 &&  \forall k\in\mathbb{Z}^2\
\text{with}\ k\notin \mathcal S.\label{(11.3.42)}
\end{align}
Notice that for any  $K=(K_1,K_2) \in \mathcal S$,  $K\not=0$, and 
\begin{equation}
\lambda_c=\frac{(\rho_K+1)(\mu\rho_K+2\alpha
)}{\rho_K}.\label{(11.3.43)}
\end{equation}
We note that for properly choosing spatial geometry, we have
\begin{align}
& \rho_K= \pi^2\left( \frac{K^2_1}{L^2_1}+\frac{K^2_2}{L^2_2}\right) = \sqrt{\frac{2\alpha}{\mu}}
\qquad \forall  K=(K_1,K_2) \in \mathcal S, 
\label{geometry}\\
& \lambda_c=2\alpha +\mu + 2 \sqrt{2\alpha \mu}.
\end{align}
 

Conditions (\ref{(11.3.41)}) and (\ref{(11.3.42)}) give rise to
a dynamic transition of (\ref{(11.3.33)}) from $(u,\lambda )=(0,\lambda_c)$.
For simplicity, we denote
$$K_1\triangleq (K_1,0),\ \ \ \ K_2\triangleq (0,K_2),$$
and $K=(K_1,K_2)\in \mathcal S$.
Also, we introduce a parameter as
\begin{align}
b = &-3\mu\rho_K+\left[12-\frac{24-15\text{sign}(K_1K_2)}{4-2\text{sign}(K_1K_2)}\right] \alpha\label{(11.3.44)}\\
&-\frac{(2\mu\rho_K+\alpha
)(2\mu\rho^2_K+28\alpha\rho_K+4\alpha
-\mu\rho_K)}{[1+\text{sign}(K_1K_2)]\cdot [(\mu\rho_{2K}+2\alpha
)(1+\rho_{2K})-\lambda_c\rho_{2K}]}\nonumber\\
&-\frac{2(2\mu\rho_K\rho_{K_1}+4\alpha\rho_{K_1}-3\alpha\rho_K)}{(1+\text{sign}K_1)\rho^2_K[(\mu\rho_{2K_1}
+2\alpha )(1+\rho_{2K_1})-\lambda_c \rho_{2K_1}]}\nonumber\\
&\ \ \times [(\mu\rho_K+2\alpha
)(2\rho^2_{K_1}-6\rho_{K_1}\rho_{K_2}-\rho_K)+6\alpha\rho_K(4\rho_{K_1}+1)]\nonumber\\
&-\frac{2(2\mu\rho_K\rho_{K_2}+4\alpha\rho_{K_2}-3\alpha\rho_K)}{(1+\text{sign}K_2)\rho^2_K((\mu\rho_{2K_2}+
2\alpha )(1+\rho_{2K_2})-\lambda_c\rho_{2K_2})}\nonumber\\
&\ \ \times [(\mu\rho_K+2\alpha
)(2\rho^2_{K_2}-6\rho_{K_1}\rho_{K_2}-\rho_K)+6\alpha\rho_K(4\rho_{K_2}+1)].\nonumber
\end{align}

The following is the main dynamic transition theorem, providing 
a precise criterion for the transition type   and the pattern formation mechanism of the system.   

\bt\la{t11.3.1}
Let $b$ be the parameter defined by
(\ref{(11.3.44)}). Assume that the eigenvalue $\beta_k$ satisfying
(\ref{(11.3.41)}) is simple. Then, for the system (\ref{(11.3.33)}) we
have the following assertions:

\begin{itemize}

\item[(1)] The system always undergoes a dynamic transition at  $(u,\lambda )=(0,\lambda_c)$. Namely, 
the basic state $u=0$ is asymptotically stable for $\lambda <\lambda_c$,
and is unstable for $\lambda >\lambda_c$.

\item[(2)] For the case where  $b<0$, this transition is continuous (Type-I). I  particular, the system 
bifurcates from $(0,\lambda_c)$ to two steady state solutions on
$\lambda >\lambda_c$, which can be expressed as
\begin{equation}
u^{\pm}_1(x,\lambda )=\pm\frac{1}{2}\sqrt{\frac{\beta_K(\lambda
)}{2|b|}}\cos\frac{K_1\pi x_1}{L_1}\cos\frac{K_2\pi
x_2}{L_2}+o\left(\beta^{{1}/{2}}_K\right),\label{(11.3.45)}
\end{equation}
and $u^{\pm}_1(x,\lambda )$ are attractors.

\item[(3)]  For the case $b>0$, this transition is jump (Type-II), and the system
has two saddle-node bifurcation solutions at some
$\lambda^*(0<\lambda^*<\lambda_c)$ such that there are 
two branches $v^{\lambda}_1$ and $v^{\lambda}_2$ of steady states 
bifurcated from $(v^*,\lambda^*)$, and there are two other branches  $v^{\lambda}_3$ and
$v^{\lambda}_4$ bifurcated from $(u^*,\lambda^*)$. In addition,
$v^{\lambda}_1$ and $v^{\lambda}_3$ are saddles, $v^{\lambda}_2$ and
$v^{\lambda}_4$ are attractors, with
$v^{\lambda}_1,v^{\lambda}_3\rightarrow 0$ as
$\lambda\rightarrow\lambda_c$.
\end{itemize}
\et

Two remarks are now in order.

\br
{\rm 
From the pattern formation point of view, for the Type-I transition, the patterns described by the transition solutions given in  (\ref{(11.3.45)}) are either lamella or rectangular:
\begin{align*}
& \text{lamella pattern} && \text{  for } K_1K_2=0, \\
& \text{rectangular pattern } && \text { for } K_1 K_2 \not=0.
\end{align*}
In the case where $b>0$,  the system undergoes a more drastic change. 
As $\lambda^\ast < \lambda < \lambda_c$,  the homogeneous state, 
the new patterns
$v_2^\lambda$ and $v_4^\lambda$ are metastable. 
For $\lambda> \lambda_c$, the system undergoes transitions to more complex patterns away from the basic homogeneous sate form.
}
\er

\br\la{r11.3.3}
{\rm
 If we take the growth term $f(u)$ as $f=\alpha
u_1(1-u_1)$ instead of $f=\alpha u_1(1-u^2_1)$ in (\ref{(11.3.29)}), (\ref{(11.3.30)}) and (\ref{(11.3.33)}), 
then  Theorem~\ref{t11.3.1} still holds true except the assertion on the existence of the two saddle-node bifurcation solutions, and the parameter should be replaced by 
\begin{align*}
b=&-\mu\rho_K+\alpha -\frac{(2\mu\rho_K+\alpha
)(2\mu\lambda^2_K+10\alpha\rho_K+\alpha
-\mu\rho_K)}{2(1+\text{sign}K_1K_2)[(\mu\rho_{2K}+\alpha
)(1+\rho_{2K})-\lambda_c\rho_{2K}]}\\
&-\frac{(2\mu\rho_K\rho_{K_1}+2\alpha\rho_{K_1}-\alpha\rho_K)}{(1+\text{sign}K_1)\lambda^2_K[(\mu\rho_{2K_1}+
\alpha )(1+\rho_{2K_1})-\lambda_c\rho_{2K_1}]}\\
&\ \ \times [(\mu\rho_K+\alpha
)(2\lambda^2_{K_1}-6\rho_{K_1}\rho_{K_2}-\rho_K)+2\alpha\rho_K(4\rho_{K_1}+1)]\\
&-\frac{(2\mu\rho_K\rho_{K_2}+2\alpha\rho_{K_2}-\alpha\rho_K)}{(1+\text{sign}K_2)\lambda^2_K[(\mu\rho_{2K_2}+
\alpha )(1+\rho_{2K_1})-\lambda_c\rho_{2K_2}]}\\
&\ \ \times [(\mu\rho_K+\alpha
)(2\lambda^2_{K_1}-6\rho_{K_1}\rho_{K_2}-\rho_K)+2\alpha\rho_K(4\rho_{K_2}+1)].
\end{align*}
}
\er

\subsection{Pattern formation and dynamic transition for the general case}
Consider the general case (\ref{(11.3.29)}).   In this case, the unknown variable becomes 
$u=(u_1, u_2)$, and the basic function spaces are then defined by 
$$
H=L^2(\Omega, \mathbb R^2), \qquad H_1 = \{ u \in H^2(\Omega, \mathbb R^2) \ | \ \frac{\partial u}{\partial n}=0 \text{ on } \Omega\}.
$$
Let $L_{\lambda}:H_1\rightarrow H$ and $G:H_1\rightarrow H$ be
defined by
\begin{equation}
\begin{aligned}
&L_{\lambda}u=\left(\begin{array}{cc} \mu\Delta -2\alpha&-\Delta\\
\lambda&\Delta -1\end{array}\right) u,\\
&G(u)=\left(\begin{array}{c} -\nabla (u_1\nabla u_2)-3\alpha
u^2_1-\alpha u^3_1\\
0\end{array}\right).
\end{aligned}
\label{(11.3.31)}
\end{equation}

The 
linearized  eigenvalue problem of (\ref{(11.3.30)}) is
\begin{equation}
L_{\lambda}\varphi =\beta\varphi ,\label{(11.3.59)}
\end{equation}
where $L_{\lambda}:H_1\rightarrow H$ is defined by (\ref{(11.3.31)}).
Let $B^{\lambda}_k$ be the matrices given by
\begin{equation}
B^{\lambda}_k=\left(\begin{array}{cc} -(\mu\rho_k+2\alpha
)&\rho_k\\
\lambda&-(\rho_k+1)
\end{array}\right),\label{(11.3.60)}
\end{equation}
where $\rho_k$ are the eigenvalues as in (\ref{(11.3.37)}). It is
easy to see that all eigenvectors $\varphi_k$ and eigenvalues
$\beta_k$ of (\ref{(11.3.59)}) can be expressed as follows
\begin{align}
&
\varphi_k=\left(\begin{array}{cc} \xi_{k1}e_k\\
\xi_{k2}e_k\end{array}\right),\label{(11.3.61)}
\\
&B^{\lambda}_k\left(\begin{array}{c} \xi_{k1}\\
\xi_{k2}\end{array}\right)=\beta_k\left(\begin{array}{c} \xi_{k1}\\
\xi_{k2}\end{array}\right),\label{(11.3.62)} \end{align} 
where
$e_k$ are  as in (\ref{(11.3.37)}), and $\beta_k$ are also the
eigenvalues of $B^{\lambda}_k$. By (\ref{(11.3.60)}), $\beta_k$ can
be expressed by
\begin{equation}
\begin{aligned}
&\beta^{\pm}_k(\lambda
)=\frac{1}{2}\left[- B \pm\sqrt{B^2-4((\rho_k+1)(\mu\rho_k+2\alpha
)-\lambda\rho_k)}\right],\\
&B=(\mu +1)\rho_k+2\alpha +1.
\end{aligned}
\label{(11.3.63)}
\end{equation}
Let $\lambda_c$ be the parameter as defined by (\ref{(11.3.40)}).
It follows from (\ref{(11.3.63)}) and (\ref{(11.3.40)}) that
\begin{align}
& 
\beta^+_K(\lambda )\left\{\begin{array}{ll} <0 &\text{ if } \lambda
<\lambda_c,\\
=0&\text{ if }\lambda =\lambda_c,\\
>0&\text{ if }\lambda >\lambda_c,
\end{array}\right.\label{(11.3.64)}
\\
&
\left\{\begin{array}{ll} \text{Re}\beta^-_k(\lambda_c)<0 &\forall
k\in\mathbb{Z}^2,\\
\text{Re}\beta^+_k(\lambda_c)<0 &\forall k\in\mathbb{Z}^2\ \text{with}\
\rho_k\neq\rho_K,
\end{array}\right.\label{(11.3.65)}
\end{align}
with $K=(K_1,K_2)$ as in (\ref{(11.3.43)}).

Then we have the following dynamic transition theorem.

\bt\la{t11.3.2}
Let $b$ be the parameter defined by
(\ref{(11.3.44)}). Assume that the eigenvalue $\beta^+_K$ satisfying
(\ref{(11.3.64)}) is simple. Then the assertions of
Theorem~\ref{t11.3.1} hold true for (\ref{(11.3.30)}),  with the expression
(\ref{(11.3.45)})  replaced by
\begin{eqnarray*}
&&u^{\pm}_{\lambda}=\pm\sqrt{a\beta^+_K(\lambda
)}\left(\begin{array}{c} \rho_K+1\\
\lambda_c\end{array}\right)\cos\frac{K_1\pi
x_1}{L_1}\cos\frac{K_2\pi x_2}{L_2}+o(|\beta^+_K|^{{1}/{2}}),\\
&&a=\frac{8(\mu\rho_K+\rho_K+2\alpha +1)}{(\rho_K+1)^3|b|}.
\end{eqnarray*}
\et

\subsection{Proof of Main Theorems}
\bp[Proof of Theorem~\ref{t11.3.1}]
Assertion (1) follows directly from the general dynamic transition theorem in Chapter 2 of \cite{ptd}. 
To prove Assertions (2) and (3), we need to reduce  (\ref{(11.3.33)}) to the center manifold 
 near $\lambda
=\lambda_c$. We note that although the underlying system is now quasilinear in this general case, the center manifold reduction holds true as well; see \cite{LSW11} for details. 

To this end, let $u=x  e_k+\Phi$, where $\Phi (x)$ the center manifold function of
(\ref{(11.3.33)}). Since  $L_{\lambda}:H_1\rightarrow H$ is
symmetric, the reduced equation is given by 
\begin{equation}
\frac{dx}{dt}=\beta_K(\lambda ) x +\frac{1}{(e_K,e_K)} (G(x
e_K+\Phi ,\lambda),e_K),\label{(11.3.46)}
\end{equation}
where $G:H_1\rightarrow H$ is defined by (\ref{(11.3.34)}), and 
\begin{equation}
(e_K,e_K)=\int_{\Omega}e^2_Kdx=\frac{2-\text{sign}(K_1K_2)}{4}|\Omega
|.\label{(11.3.47)}
\end{equation}

It is known that the center manifold function satisfies that $\Phi (x)=O(x^2).$
A direct computation shows that 
\begin{eqnarray}
&&<G(xe_K+\Phi ,\lambda_c),e_K>\label{(11.3.48)}\\
&=&-\alpha x^3\int_{\Omega}e^4_Kdx-6\alpha x\int_{\Omega}e^2_K\Phi
dx\nonumber\\
&&+\lambda_cx\int_{\Omega}[e_K\nabla e_k\cdot\nabla (-\Delta
+I)^{-1}\Phi +\Phi\nabla e_k\cdot\nabla (-\Delta
+I)^{-1}e_K]dx+o(x^3).\nonumber
\end{eqnarray}
It is clear that
$$(-\Delta +I)^{-1}e_K=\frac{1}{\rho_K+1}e_K,\ \ \ \ \Delta
e_K=-\rho_Ke_K.$$ 
We infer  from (\ref{(11.3.48)}) that 
\begin{align}
(G(xe_K+\Phi ,\lambda_c),e_K )\label{(11.3.49)}
=&-\alpha x^3\int_{\Omega}e^4_Kdx-6\alpha x\int_{\Omega}e^2_K\Phi
dx \\
&+\lambda_c x\int_{\Omega}\left[\frac{1}{\rho_K+1}|\nabla
e_K|^2\Phi -|\nabla e_K|^2(-\Delta _I)^{-1}\Phi\right.\nonumber\\
&\ \ \ \ \left.+\rho_Ke^2_K(-\Delta
+I)^{-1}\Phi\right]dx+o(x^3).\nonumber
\end{align}

Using the approximation formula for center manifold  functions given in (A.11) in \cite{MW09c}, 
$\Phi$  satisfies the equation
\begin{eqnarray}
-L_{\lambda_c}\Phi &=&G_2(xe_K,\lambda_c)+o(x^2)\label{(11.3.50)}\\
&=&x^2\left[\left(\frac{\rho_K\lambda_c}{\rho_K+1}-3\alpha\right)e^2_K-\frac{\lambda_c}{\rho_K+1}|\nabla
e_K|^2\right]+o(x^2).\nonumber
\end{eqnarray}
In view of (\ref{(11.3.37)}), we find
\begin{equation}
\left.\begin{aligned}
&e^2_K=\frac{1}{4}[e_0+e_{2K_1}+e_{2K_2}+e_{2K}], \\
&|\nabla
e_K|^2=\frac{1}{4}[\rho_Ke_0+(\rho_{K_2}-\rho_{K_1})e_{2K_1}+(\rho_{K_1}-\rho_{K_2})e_{2K_2}-\rho_Ke_{2K}].
\end{aligned}\right.\label{(11.3.51)}
\end{equation}
Thus, (\ref{(11.3.50)}) is written as
\begin{align}
-L_{\lambda_c}\Phi =&\frac{x^2}{4}\left[-3\alpha
e_0+\left(\frac{2\rho_{K_1}\lambda_c}{\rho_K+1}-3\alpha\right)e_{2K_1}\right. \label{(11.3.52)}\\
& +\left(\frac{2\rho_{K_2}\lambda_c}{\rho_K+1}
-3\alpha\right)e_{2K_2}  \left.+\left(\frac{2\rho_K\lambda_c}{\rho_K+1}e_{2K}-3\alpha\right)e_{2K}\right]+o(x^2).\nonumber
\end{align}
Denote by
\begin{equation}
\Phi
=\Phi_0e_0+\Phi_{2K_1}e_{2K_1}+\Phi_{2K_2}e_{2K_2}+\Phi_{2K}e_{2K}.\label{(11.3.53)}
\end{equation}
Note that
\begin{equation}
-L_{\lambda_c}e_{2K}=\frac{1}{1+\rho_{2K}}\left[(1+\rho_{2K})(\mu\rho_K+2\alpha
)-\lambda_c\rho_{2K}\right]e_{2K}.\label{(11.3.54)}
\end{equation}
Then, by (\ref{(11.3.43)}) and (\ref{(11.3.52)})-(\ref{(11.3.54)}) we
obtain
\begin{equation}
\left.\begin{aligned} &\Phi_0=-\frac{3}{8},\\
&\Phi_{2K_1}=\frac{(1+\rho_{2K_1})(2\mu\rho_K\rho_{K_1}+4\alpha\rho_{K_1}-3\alpha\rho_K)}{4\rho_K[(1+
\rho_{2K_1})(\mu\rho_{2K_1}+2\alpha
)-\rho_{2K_1}\lambda_c]},\\
&\Phi_{2K_2}=\frac{(1+\rho_{2K_2})(2\mu\rho_K\rho_{K_2}+4\alpha\rho_{K_2}-3\alpha\rho_K)}{4\rho_K[(1+
\rho_{2K_2})(\mu\rho_{2K_2}+2\alpha
)-\rho_{2K_2}\lambda_c]},\\
&\Phi_{2K}=\frac{(1+\rho_{2K})(2\mu\rho_K+\alpha
)}{4[(1+\rho_{2K})(\mu\rho_{2K}+2\alpha
)-\rho_{2K}\lambda_c]}.
\end{aligned}
\right.\label{(11.3.55)}
\end{equation}
Inserting (\ref{(11.3.53)}) and (\ref{(11.3.37)}) into
(\ref{(11.3.49)}) we get
\begin{align}
&<G(xe_K+\Phi ,\lambda_c),e_K>\label{(11.3.56)}
= -\alpha x^3\int_{\Omega}e^4_Kdx  \\
&-\frac{6\alpha
x(2-\text{sign}(K_1K_2))}{4}\int_{\Omega}\left[\Phi_0e^2_0+\Phi_{2K_1}e^2_{2K_1}+\Phi_{2K_2}e^2_{2K_2}+\Phi_{2K}e^2_{2K}\right]dx\nonumber\\
&+\frac{\lambda_cx(2-\text{sign}(K_1K_2))}{4(\rho_K+1)}\int_{\Omega}\left[\rho_K\Phi_0e^2_0+(\rho_{K_2}-\rho_{K_1})
\Phi_{2K_1}e^2_{2K_1}\right.\nonumber\\
&\ \ \ \ \ \ \ \
\left.+(\rho_{K_1}-\rho_{K_2})\Phi_{2K_2}e^2_{2K_2}-\rho_K\Phi_{2K}e^2_{2K}\right]dx\nonumber\\
&-\frac{\lambda_cx(2-\text{sign}(K_1K_2))}{4}\int_{\Omega}\left[\rho_K\Phi_0e^2_0+\frac{\rho_{K_2}-\rho_{K_1}}{1+
\rho_{2K_1}}\Phi_{2K_1}e^2_{2K_1}\right.\nonumber\\
&\ \ \ \ \ \ \ \
\left.+\frac{\rho_{K_1}-\rho_{K_2}}{1+\rho_{2K_2}}\Phi_{2K_2}e^2_{2K_2}-\frac{\rho_K}{1+\rho_{2K}}\Phi_{2K}
e^2_{2K}\right]dx\nonumber\\
&+\frac{\lambda_c\rho_Kx(2-\text{sign}(K_1K_2))}{4}\int_{\Omega}\left[\Phi_0e^2_0+\frac{\Phi_{2K_1}}{1+\rho_{2K_1}}
e^2_{2K_1}+\frac{\Phi_{2K_2}}{1+\rho_{2K_2}}e^2_{2K_2}\right.\nonumber
\\
&\ \ \ \ \ \ \ \
\left.+\frac{\Phi_{2K}}{1+\rho_{2K}}e^2_{2K}\right]dx+o(x^3)\nonumber\\
&=-\alpha x^3\int_{\Omega}e^4_Kdx+\frac{|\Omega
|x(2-\text{sign}(K_1K_2))}{4}\nonumber\\
&\times\left[(\mu\rho_K-4\alpha
)\Phi_0+\frac{1}{1+\text{sign}K_1}\left(\frac{\lambda_c(\rho_{K_2}-\rho_{K_1})}{1+\rho_K}+\frac{2\lambda_c\rho_{K_1}}
{1+\rho_{2K_1}}-6\alpha\right)\Phi_{2K_1}\right.\nonumber\\
&\ \
+\frac{1}{1+\text{sign}K_2}\left(\frac{\lambda_c(\rho_{K_1}-\rho_{K_2})}{1+\rho_K}+\frac{2\lambda_c\rho_{K_2}}
{1+\rho_{2K_1}}-6\alpha\right)\Phi_{2K_2}\nonumber\\
&\ \
\left.+\frac{1}{2(1+\text{sign}(K_1K_2))}\left(-\frac{\lambda_c\rho_K}{1+\rho_K}+\frac{2\lambda_c\rho_K}{1+\rho_{2K}}
-6\alpha\right)\Phi_{2K}\right]dx+o(x^3).\nonumber
\end{align}
Also, we note  that
$$\int_{\Omega}e^4_K=\int^{L_1}_0 e^4_{K_1}dx_1\int^{L_2}_0e^4_{K_2}dx_2=\frac{24-15\text{sign}(K_1K_2)}{64}.$$
Then, putting (\ref{(11.3.55)}) into (\ref{(11.3.56)}) we get
\begin{equation}
<G(xe_K+\Phi ,\lambda_c),e_K>=\frac{(2-\text{sign}(K_1K_2))|\Omega
|x^3}{32}b+o(x^3),\label{(11.3.57)}
\end{equation}
where $b$ is the parameter given by (\ref{(11.3.44)}).

By  (\ref{(11.3.46)}) and (\ref{(11.3.57)}), we derive the reduced equation on the center manifold as follows:
\begin{equation}
\frac{dx}{dt}=\beta_K(\lambda)x+\frac{b}{8}x^3+o(x^3).\label{(11.3.58)}
\end{equation}
Based on the dynamic transition theory developed in Chapter 2 in \cite{ptd}, we obtain 
Assertions (2) and (3), except that two saddle-node bifurcations
occur at the same point $\lambda =\lambda^*$. To prove this
conclusion, we note that if $u^*(x)$ is a steady state solution of
(\ref{(11.3.33)}), then
$$v^*(x)=u^*(x+\pi )=u^*(x-\pi )$$
is also a steady state solution of (\ref{(11.3.33)}). This is because
the eigenvectors (\ref{(11.3.37)}) form an orthogonal base of $H_1$. 
Hence, two saddle-node bifurcations on $\lambda <\lambda_c$ imply
that they must occur at the same point $\lambda =\lambda^*$. Thus the proof of  the theorem is complete.
\ep

\bp[Proof of Theorem~\ref{t11.3.2}]
 Assertion (1) follows from (\ref{(11.3.64)}) and
(\ref{(11.3.65)}). To prove Assertions (2) and (3), we need to get
the reduced equation of (\ref{(11.3.30)}) to the center manifold near
$\lambda =\lambda_c$.

Let $u=x\cdot\varphi_K+\Phi$, where $\varphi_K$ is the eigenvector
of (\ref{(11.3.59)}) corresponding to $\beta_K$ at $\lambda
=\lambda_c$, and $\Phi (x)$ the center manifold function of
(\ref{(11.3.30)}). Then the reduced equation of (\ref{(11.3.30)}) read
\begin{equation}
\frac{dx}{dt}=\beta^+_K(\lambda)x+\frac{1}{<\varphi_K,\varphi^*_K>}<G(x\cdot\varphi_K+\Phi
),\varphi^*_K>,\label{(11.3.66)}
\end{equation}
Here $\varphi^*_K$ is the conjugate eigenvector of $\varphi_K$.

By (\ref{(11.3.61)}) and (\ref{(11.3.62)}), $\varphi_K$ is written as
\begin{equation}
\varphi_K=(\xi_1e_k,\ \  \xi_2e_K)^T,\label{(11.3.67)}
\end{equation}
with $(\xi_1,\xi_2)$ satisfying
\begin{equation}
\left(\begin{array}{cc}
-(\mu\rho_K+2\alpha )&\rho_K\\
\lambda_c&-(\rho_K+1)\end{array}\right)\left(\begin{array}{c}
\xi_1\\
\xi_2\end{array}\right)=0,\label{(11.3.68)}
\end{equation}
from which we get
\begin{equation}
(\xi_1,\xi_2)=(\rho_K+1,\ \ \lambda_c).\label{(11.3.69)}
\end{equation}
Likewise, $\varphi^*_K$ is
\begin{equation}
\varphi^*_K=(\xi^*_1e_k,\ \ \xi^*_2e_k)^T,\label{(11.3.70)}
\end{equation}
with $(\xi^*_1,\xi^*_2)$ satisfying
$$\left(\begin{array}{cc}
-(\mu\rho_K+2\alpha )&\lambda_c\\
\rho_K&-(\rho_K+1)\end{array}\right)\left(\begin{array}{c}
\xi^*_1\\
\xi^*_2\end{array}\right)=0,$$ which yields
\begin{equation}
(\xi^*_1,\xi^*_2)=(\rho_{K+1},\rho_K).\label{(11.3.71)}
\end{equation}

By (\ref{(11.3.31)}), the nonlinear operator $G$ is
\begin{eqnarray*}
&&G(u_1,u_2)=G_2(u_1,u_2)+G_3(u_1,u_2),\\
&&G_2(u_1,u_2)=-(\nabla u_1\nabla u_2+u_1\Delta u_2+3\alpha
u^2_1)\left(\begin{array}{c} 1\\
0\end{array}\right),\\
&&G_3(u_1,u_2)=-\alpha u^3\left(\begin{array}{c} 1\\
0\end{array}\right).
\end{eqnarray*}
It is known that the center manifold function
$$\Phi (x)=(\Phi_1(x),\Phi_2(x))=O(x^2).$$
Then, in view of (\ref{(11.3.67)}) and (\ref{(11.3.69)}), by direct
computation we derive that
\begin{align}
& (G(x\xi_1e_K+\Phi_1,\ \ x\xi_2e_K+\Phi_2),\ \
\varphi^*_K )\label{(11.3.72)}\\
&= ( xG_2(\xi_1e_K,\Phi_2)+ x G_2(\Phi_1,\xi_2e_K) +x^3 G_3(\xi_1e_K,\xi_2e_K),\varphi^*_K) +o(x^3)\nonumber\\
&=x\xi^*_1\int_{\Omega}[\xi_2\Phi_1|\nabla
e_K|^2-\frac{1}{2}\xi_1\Delta\Phi_2e^2_K-6\alpha\xi_1\Phi_1e^2_K]dx\nonumber\\
&\quad -\alpha\xi^*_1\xi^3_1x^3\int_{\Omega}e^4_Kdx+o(x^3).\nonumber
\end{align}

Using the approximation formula for center manifold  functions given in (A.11) in \cite{MW09c}, $\Phi
=(\Phi_1,\Phi_2)$ satisfies
\begin{eqnarray}
-L_{\lambda_c}\Phi&=&-x^2G_2(\xi_1e_k,\xi_2e_K)+o(x^2)\label{(11.3.73)}\\
&=&-x^2(\xi_1\xi_2|\nabla
e_K|^2+(3\alpha\xi^2_1-\xi_1\xi_2\rho_K)e^2_K)\left(\begin{array}{c}
1\\
0\end{array}\right)+o(x^2).\nonumber
\end{eqnarray}
From (\ref{(11.3.37)}) we see that
\begin{align*}
e^2_K=&\frac{1}{4}(1+e_{2K_1})(1+e_{2K_2})=\frac{1}{4}(e_0+e_{2K_1}+e_{2K_2}+e_{2K}),
\\
|\nabla
e_K|^2=&\frac{\rho_{K_1}}{4}(1-e_{2K_1})(1+e_{2K_2})+\frac{\rho_{K_2}}{4}(1+e_{2K_1})(1-e_{2K_2})\\
=&\frac{\rho_K}{4}e_0+\frac{\rho_{K_2}-\rho_{K_1}}{4}e_{2K_1}+\frac{\rho_{K_1}-\rho_{K_2}}{4}e_{2K_2}-
\frac{\rho_K}{4}e_{2K}.
\end{align*}
Thus, (\ref{(11.3.73)}) is written as
\begin{align}
L_{\lambda_c}\Phi
=&-\frac{\xi_1x^2}{4}(3\alpha\xi_1e_0+(3\alpha\xi_1-2\xi_2\rho_{K_1})e_{2K_1}\label{(11.3.74)}\\
&+(3\alpha\xi_1-2\xi_2\rho_{K_2})e_{2K_2}+(3\alpha\xi_1-2\xi_2\rho_K)e_{2K})\left(\begin{array}{c}
1\\
0\end{array}\right) +o(x^3).\nonumber
\end{align}
Let
\begin{equation}
\left(\begin{array}{c} \Phi_1\\
\Phi_2\end{array}\right)=\left(\begin{array}{c} \Phi^0_1\\
\Phi^0_2\end{array}\right)e_0+\left(\begin{array}{c}
\Phi^{2K_1}_1\\
\Phi^{2K_1}_2\end{array}\right)e_{2K_1}+\left(\begin{array}{c}
\Phi^{2K_2}_1\\
\Phi^{2K_2}_2\end{array}\right)e_{2K_2}+\left(\begin{array}{c}
\Phi^{2K}_1\\
\Phi^{2K}_2\end{array}\right)e_{2K}\label{(11.3.75)}
\end{equation}
It is clear that
$$L_{\lambda}\left(\begin{array}{c}
\Phi^k_1\\
\Phi^k_2\end{array}\right)e_k=B^{\lambda}_k\left(\begin{array}{c}
\Phi^k_1\\
\Phi^k_2\end{array}\right)e_k,$$ where $B^{\lambda}_k$ is the matrix
given by (\ref{(11.3.60)}). Then by  (\ref{(11.3.74)})
and (\ref{(11.3.75)}) we have
$$\left(\begin{array}{c}
\Phi^{2k}_1\\
\Phi^{2k}_2\end{array}\right)=-\frac{(3\alpha\xi^2_1-2\xi_1\xi_2\rho_k)  x^2}{4}B^{-1}_{2k}\left(\begin{array}{c}
1\\
0\end{array}\right),$$ for $k=K,K_1,K_2$, and
$B_{2k}=B^{\lambda_c}_{2k}$.

Direct computation shows that
\begin{eqnarray}
&&\left(\begin{array}{c} \Phi^0_1\\
\Phi^0_2\end{array}\right)=\frac{3\xi^2_1x^2}{8}\left(\begin{array}{c}
1\\
\lambda_c\end{array}\right),\label{(11.3.76)}\\
\nonumber\\
&&\left(\begin{array}{c} \Phi^{2K_1}_1\\
\Phi^{2K_2}_2
\end{array}\right)=\frac{\xi_1(3\alpha\xi_1-2\xi_2\rho_{K_1})}{4\text{det}B_{2K_1}}\left(\begin{array}{c}
1+\rho_{2K_1}\\
\lambda_c\end{array}\right),\label{(11.3.77)}\\
\nonumber\\
&&\left(\begin{array}{c} \Phi^{2K_2}_1\\
\Phi^{2K_2}_2\end{array}\right)=\frac{\xi_1(3\alpha\xi_1-2\xi_2\rho_{K_2})}{4\text{det}B_{2K_2}}\left(\begin{array}{c}
1+\rho_{2K_2}\\
\lambda_c\end{array}\right),\label{(11.3.78)}\\
\nonumber\\
&&\left(\begin{array}{c} \Phi^{2K}_1\\
\Phi^{2K}_2\end{array}\right)=\frac{\xi_1(3\alpha\xi_1-2\xi_2\rho_K)}{4\text{det}B_{2K}}\left(\begin{array}{c}
1+\rho_{2K}\\
\lambda_c\end{array}\right).\label{(11.3.79)}
\end{eqnarray}
Inserting (\ref{(11.3.76)}) into (\ref{(11.3.72)}), by
(\ref{(11.3.69)}) and (\ref{(11.3.71)}) we get
\begin{eqnarray*}
&&<G(x\varphi_K+\Phi
),\varphi^*_K>=\frac{(2-\text{sign}K_1K_2)(\rho_K+1)|\Omega
|}{8}\\
&&\times\left[-\frac{8\alpha
(\rho_K+1)^3x^3\int_{\Omega}e^4_Kdx}{(2-\text{sign}K_1K_2)|\Omega
|}+2(\xi_2\rho_K-6\alpha\xi_1)\Phi^0_1x\right.\\
&&\ \
+\frac{2}{1+\text{sign}K_1}(\xi_2(\rho_{K_2}-\rho_{K_1})-6\alpha\xi_1)\Phi^{2K_1}_1x\\
&&\ \
+\frac{2}{1+\text{sign}K_2}(\xi_2(\rho_{K_1}-\rho_{K_2})-6\alpha\xi_1)\Phi^{2k_2}_2x\\
&&\ \
-\frac{2-\text{sign}K_1K_2}{2}(\xi_2\rho_K+6\alpha\xi_1)\Phi^{2K}_1x\\
&&\ \
+\frac{(\rho_K+1)\rho_{2K_1}}{1+\text{sign}K_1}\Phi^{2K_1}_2x
+\frac{(\rho_K+1)\rho_{K_2}}{1+\text{sign}K_2}\Phi^{2K_2}_2x\\
&&\ \
\left.+\frac{(\rho_K+1)\rho_{2K}(2-\text{sign}K_1K_2)}{4}\Phi^{2K}_2x\right]+o(x^3).
\end{eqnarray*}
By definition, we have
\begin{align*}
& \rho_{K_1}+\rho_{K_2}=\rho_K,\ \ \ \
\rho_{2K}=4\rho_K \qquad \qquad  \forall K=(K_1,K_2),\\
& <\varphi,\varphi^*>= \left[(\rho_K+1)^2+\rho_K\lambda_c\right]\int_\Omega e^2_Kdx\\
& \qquad \qquad =\frac{2-\text{sign}(K_1K_2)}{4}(\rho_K+1)(\mu\rho_K+\rho_K+2\alpha
+1)|\Omega |. 
\end{align*}

In view of (\ref{(11.3.76)})-(\ref{(11.3.79)}),
the reduced equation (\ref{(11.3.66)}) is given by 
\begin{equation}
\frac{dx}{dt}=\beta^+_K(\lambda)x+\frac{(\rho_K+1)^3bx^3}{8(\mu\rho_K+\rho_K+2\alpha
+1)}+o(x^3),\label{(11.3.81)}
\end{equation}
where $b$ is the parameter as in (\ref{(11.3.44)}). Then the theorem follows readily from (\ref{(11.3.81)}). 
The proof is complete.
\ep

\section{Transition of Three-Component Systems}
\subsection{The model}
Hereafter $\delta_0\geq 0$ is always assumed to be  a constant. Hence,
(\ref{(11.3.22)}) has a positive constant steady state $u^*$ given by
\begin{equation}
(u^*_1,u^*_2,u^*_3) \quad \text{ with } 
u^*_1=\left(\frac{u^*_3}{1+u^*_3}\right)^{1/2},\ \ \ \ u^*_2=\lambda u^*_1,\ \ \ \
u^*_3u^*_1=\frac{\delta_0}{\delta}.
\label{(11.3.23)}
\end{equation}
It is easy to see that $u^*_3$ is the unique positive real root of the cubic equation
$$x^3-\left(\frac{\delta_0}{\delta}\right)^2x-\left(\frac{\delta_0}{\delta}\right)^2=0.$$
 Consider  the translation
\begin{equation}
(u_1, u_2, u_3) \rightarrow (u_1^*+u_1, u_2^*+u_2, u_1^*+u_1).\label{(11.3.24)}
\end{equation}
Then  equations (\ref{(11.3.22)}) are equivalent to 
\begin{equation}
\begin{aligned}
&\frac{\partial u_1}{\partial t}=\mu\Delta u_1-2\alpha
u^{*2}_1u_1-u^*_1\Delta u_2+\frac{\alpha
u^*_1}{(1+u^*_3)^2}u_3+g(u),\\
&\frac{\partial u_3}{\partial t}=r\Delta u_3-\delta u^*_1u_3-\delta
u^*_3u_1-\delta u_1u_3,\\
&-\Delta u_2+u_2=\lambda u_1,\\
&\left.\frac{\partial (u_1, u_2, u_3)}{\partial n}\right|_{\partial\Omega}=0,\\
&u(0)=u_0,
\end{aligned}
\label{(11.3.25)}
\end{equation}
where $u=(u_1, u_3)$, $u_2=\lambda [-\triangle + 1]^{-1}u_1$, and 
\begin{eqnarray}
g(u)&=&-\nabla (u_1\nabla u_2)-3\alpha u^*_1u^2_1-\alpha
u^3_1+\frac{\alpha
(u_1+u^*_1)(u_3+u^*_3)}{1+u^*_3+u_3}\label{(11.3.26)}\\
&&-\frac{\alpha u^*_1u^*_3}{1+u^*_3}-\frac{\alpha
u^*_1u_3}{(1+u^*_3)^2}-\frac{\alpha u^*_3u_1}{1+u^*_3}.\nonumber
\end{eqnarray}
The Taylor expansion of $g$ at $u=0$ is expressed by
\begin{eqnarray*}
g(u)&=&-\nabla (u_1\nabla u_2)-3\alpha u^*_1u^2_1+\frac{\alpha
u_1u_3}{(1+u^*_3)^2}-\frac{\alpha u^*_1u^2_3}{(1+u^*_3)^3}\\
&&-\alpha u^3_1-\frac{\alpha u_1u^2_3}{(1+u^*_3)^3}+\frac{\alpha
u^*_1u^3_3}{(1+u^*_3)^4}+o(3).
\end{eqnarray*}
Let
\begin{eqnarray*}
&&H=L^2(\Omega ,\R^2),\\
&&H_1=\{u\in H^2(\Omega ,\R^2)|\ \frac{\partial u}{\partial n}=0\
\text{on}\ \partial\Omega\}.
\end{eqnarray*}
Define the operators $L_{\lambda}:H_1\rightarrow H$
and $G_{\lambda}:H_1\rightarrow H$ by
\begin{equation}
\begin{aligned}
&L_{\lambda}u=\left(\begin{array}{cc} \mu\Delta -2\alpha
u^{*2}_1-\lambda u^*_1\Delta [-\Delta +I]^{-1}&\frac{\alpha
u^*_1}{(1+u^*_3)^2}\\
-\delta u^*_3&r\Delta -\delta
u^*_1\end{array}\right)\left(\begin{array}{c} u_1\\
u_3\end{array}\right),\\
&G(u,\lambda )=\left(\begin{array}{c} g(u)\\
-\delta u_1u_3\end{array}\right),
\end{aligned}
\label{(11.3.27)}
\end{equation}
Then the problem
(\ref{(11.3.25)}) takes the following  the abstract form:
\begin{equation}
\begin{aligned}
&\frac{du}{dt}=L_{\lambda}u+G(u,\lambda ),\\
&u(0)=u_0.
\end{aligned}
\label{(11.3.28)}
\end{equation}

It is known that the inverse mapping
$$[-\Delta +I]^{-1}:H\rightarrow H_1$$
is a bounded linear operator. Therefore we have
\begin{eqnarray*}
&&L_{\lambda}:H_1\rightarrow H\ \text{is\ a\ sector\ operator,\ and}\\
&&G_{\lambda}:H_{\theta}\rightarrow H\ \text{is}\ C^{\infty}\
\text{bounded\ operator\ for}\ \theta\geq\frac{1}{2}.
\end{eqnarray*}
We note that the transition of (\ref{(11.3.25)}) from $u=0$ is equivalent
to that of (\ref{(11.3.22)}) from $u=u^*$.

Theorems~\ref{t11.3.1} and \ref{t11.3.2} show that a two-component system undergoes only a dynamic transition to 
steady states. As we shall see, the transition for the three-component  system (\ref{(11.3.21)}) is
quite different -- it can undergo  both steady state and spatiotemporal transitions.  

\subsection{Linearized eigenvalue of (\ref{(11.3.22)})}

The eigenvalue equations of (\ref{(11.3.22)}) at the steady state
$(u_1^*, u_2^\ast, u_3^\ast)$ given by (\ref{(11.3.23)}) in their 
abstract form are given by 
\begin{equation}
L_{\lambda}\varphi =\beta\varphi ,\label{(11.3.82)}
\end{equation}
where $L_{\lambda}:H_1\rightarrow H$ as defined in (\ref{(11.3.27)}).
The explicit form of (\ref{(11.3.82)}) is given by 
$$\left(\begin{array}{cc}
\mu\Delta -2\alpha u^{*2}_1-\lambda u^*_1\Delta [-\Delta
+I]^{-1}&\frac{\alpha u^*_1}{(1+u^*_3)^2}\\
-\delta u^*_3&r\Delta -\delta
u^*_1\end{array}\right)\left(\begin{array}{c} \psi_1\\
\psi_3\end{array}\right)=\beta\left(\begin{array}{c} \psi_1\\
\psi_3\end{array}\right).$$ 
As before, let $\rho_k$ and $e_k$ be the
eigenvalue and eigenvector of  $-\triangle$ with Neumann boundary condition given by  (\ref{(11.3.37)}), and let 
$$\psi_k=(\psi^k_1,\psi^k_3)=(\xi_{k1}e_k,\xi_{k3}e_k).$$
Then, it is easy to see that $\psi_k$ is an eigenvector of
(\ref{(11.3.82)}) provided that $(\xi_{k1},\xi_{k3})\in \R^2$ is an
eigenvector of the matrix $A^{\lambda}_k$:
$$A^{\lambda}_k\left(\begin{array}{c}
\xi_{k1}\\
\xi_{k3}\end{array}\right)=\beta_k\left(\begin{array}{c} \xi_{k1}\\
\xi_{k3}\end{array}\right),
$$ 
with
\begin{equation}
A^{\lambda}_k=\left(\begin{array}{cc}
\frac{\lambda\rho_k u^*_1}{1+\rho_k}-\mu\rho_k-2\alpha
u^{*2}_1&\frac{\alpha u^*_1}{(1+u^*_3)^2}\\
-\delta u^*_3&-r\rho_k-\delta
u^*_1\end{array}\right).\label{(11.3.83)}
\end{equation}
The eigenvalues $\beta_k$ of $A^{\lambda}_k$, which are also
eigenvalues of (\ref{(11.3.82)}), are expressed by
\begin{equation}
\begin{aligned}
&\beta^{\pm}_k(\lambda
)=\frac{1}{2}\left[a\pm\sqrt{a^2-4\text{det}A^{\lambda}_k}\right],\\
&a=\text{tr}A^{\lambda}_k=\frac{\lambda\rho_k u^*_1}{1+\rho_k}-\mu\rho_k-2\alpha
u^{*2}_1-r\rho_k-\delta u^*_1.
\end{aligned}\label{(11.3.84)}
\end{equation}

To derive the PES, we introduce two parameters as follows:
\begin{align}
& 
\Lambda_c=\min_{\rho_K}\frac{(\rho_K+1)}{\rho_Ku^*_1}\left[\mu\rho_K+2\alpha
u^{*2}_1+r\rho_K+\delta u^*_1\right],\label{(11.3.85)}
\\
&
\lambda_c=\min_{\rho_K}\frac{(\rho_K+1)}{\rho_Ku^*_1}\left[\mu\rho_K+2\alpha
u^{*2}_1+\frac{\alpha\delta}{(1+u^*_3)^2(r\rho_K+\delta
u^*_1)}\right].\label{(11.3.86)}
\end{align}
Let $K=(K_1,K_2)$ and $K^*=(K^*_1,K^*_2)$ be the integer pairs  such  that
$\rho_K$ and $\rho_{K^*}$ satisfy (\ref{(11.3.85)}) and
(\ref{(11.3.86)}) respectively.

\bt\la{t11.3.3}
Let $\Lambda_c$ and $\lambda_c$ be the
parameters defined by (\ref{(11.3.85)}) and (\ref{(11.3.86)})
respectively. Then we have the following assertions:

\begin{itemize}

\item[(1)] As $\Lambda_c<\lambda_c$, the eigenvalues $\beta^{\pm}_K(\lambda
)$ of (\ref{(11.3.84)}) are a pair of conjugate complex numbers near
$\lambda =\Lambda_c$, and all eigenvalues of (\ref{(11.3.84)})
satisfy
\begin{align}
&
\text{Re}\ \beta^{\pm}_K(\lambda )\left\{\begin{array}{ll} <0&\lambda
<\Lambda_c,\\
=0&\lambda =\Lambda_c,\\
>0&\lambda >\Lambda_c,
\end{array}\right.\label{(11.3.87)}
\\
&
\text{Re}\ \beta^{\pm}_k(\Lambda_c)<0,\ \ \ \ \forall
k\in\mathbb{Z}^2\ \text{with}\ \rho_k\neq\rho_K\label{(11.3.88)}
\end{align}

\item[(2)] As $\lambda_c<\Lambda_c$, the eigenvalue $\beta^+_{K^*}(\lambda
)$ is real near $\lambda =\lambda_c$, and all of (\ref{(11.3.84)})
satisfy
\begin{align}
&
\beta^+_{K^*}(\lambda )\left\{\begin{array}{ll} <0,&\lambda
<\lambda_c,\\
=0,&\lambda =\lambda_c,\\
>0,&\lambda >\lambda_c,
\end{array}\right.\label{(11.3.89)}
\\
&
\left\{\begin{array}{ll} \text{Re} \beta^+_k(\lambda_c)<0,&\forall
k\in\mathbb{Z}^2\ \text{with}\ \rho_k\neq\rho_{K^\ast},\\
\text{Re} \beta^-_k(\lambda_c)<0,&\forall |k|\geq 0.
\end{array}\right.\label{(11.3.90)}
\end{align}
\end{itemize}
\et

\bp
By (\ref{(11.3.84)}) we can see that
$\beta^{\pm}_k(\lambda )$ are a pair of complex eigenvalues of
(\ref{(11.3.82)}) near some $\lambda =\lambda^*$, and satisfy
$$\text{Re} \beta^{\pm}_k(\lambda )\left\{\begin{array}{ll}
<0,&\lambda <\lambda^*,\\
=0,&\lambda =\lambda^*,\\
>0,&\lambda >\lambda^*,
\end{array}\right.$$
if and only if
$$\text{tr} A^{\lambda^*}_k=0,\ \ \ \ \text{det}\ A^{\lambda^*}_k>0.$$
Likewise, $\beta^+_k(\lambda )$ is real near $\lambda =\lambda^*$
and satisfies
$$\beta^+_k(\lambda )\left\{\begin{array}{ll}
<0,&\lambda <\lambda^*,\\
=0,&\lambda =\lambda^*,\\
>0,&\lambda >\lambda^*,
\end{array}\right.$$
if and only if
$$\text{tr} A^{\lambda^*}_k<0,\ \ \ \ \text{det}\ A^{\lambda^*}_k=0$$
Due to the definition of $\lambda_c$ and $\Lambda_c$, when
$\Lambda_c<\lambda_c$ we have
\begin{equation}
\begin{aligned}
&\text{tr}\ A^{\Lambda_c}_K=0,\\
&\text{tr}\ A^{\Lambda_c}_k<0,\ \ \ \ \forall k\in\mathbb{Z}^2\
\text{with}\ \rho_k\neq\rho_K,
\end{aligned}
\label{(11.3.91)}
\end{equation}
$$\text{det}\ A^{\Lambda_c}_k>0,\ \ \ \ \forall |k|\geq 0,$$
and when $\lambda_c<\Lambda_c$,
\begin{equation}
\begin{aligned}
&\text{det}\ A^{\lambda_c}_{K^*}=0,\\
&\text{det}\ A^{\lambda_c}_k>0,\ \ \ \ \forall k\in\mathbb{Z}^2\
\text{with}\ \rho_k\neq\rho_K,\\
&\text{tr}\ A^{\lambda_c}_k<0,\ \ \ \ \forall |k|\geq 0.
\end{aligned}
\label{(11.3.92)}
\end{equation}
It is known that the real parts of $\beta^{\pm}_k$ are negative at
$\lambda$ if and only if
$$\text{det}\ A^{\lambda}_k>0,\ \ \ \ \text{tr}\ A^{\lambda}_k<0.$$
Hence, Assertions (1) and (2) follow from (\ref{(11.3.91)}) and
(\ref{(11.3.92)}) respectively. The theorem is proved.
\ep

\subsection{Dynamic transition theorem for (\ref{(11.3.22)})}

Based on Theorem~\ref{t11.3.3}, we immediately get the following transition
theorem for (\ref{(11.3.25)}).

\bt\la{t11.3.4} 
Let $\Lambda_c$ and $\lambda_c$ be given by
(\ref{(11.3.85)}) and (\ref{(11.3.86)}) respectively. Then the
following assertions hold true for (\ref{(11.3.25)}):

\begin{itemize}

\item[(1)] When $\Lambda_c<\lambda_c$, the system undergoes a dynamic  transition to
periodic solutions at $(u,\lambda )=(0,\Lambda_c)$. In particular,
if the eigenvalues $\beta^{\pm}_K$ satisfying (\ref{(11.3.87)}) are
complex simple, then there is a parameter $b_0$ such that the dynamic 
transition is continuous (Type-I) as $b_0<0$, and is jump (Type-II)
as $b_0>0$ with a singularity separation of periodic solutions at
some $\lambda^*<\Lambda_c$.

\item[(2)] When $\lambda_c<\Lambda_c$, the system undergoes   a dynamic transition to
steady states at $(u,\lambda )=(0,\lambda_c)$. If
$\beta^+_{K^*}(\lambda )$ satisfying (\ref{(11.3.89)}) is simple,
then there exists a parameter $b_1$ such that the transition is
continuous as $b_1<0$, and jumping as $b_1>0$ with two saddle-node
bifurcations at some $\tilde{\lambda}<\lambda_c$ from
$(u^+,\tilde{\lambda})$ and $(u^-,\tilde{\lambda})$.
\end{itemize}
\et

\br\la{r11.3.5}
{\rm 
By applying the standard procedure used in the
preceding sections, we can derive explicit formulas for the two
parameters $b_0$ and $b_1$ in Theorem~\ref{t11.3.4}. However, due to their complexity, we omit the details. 
Instead in the following, we shall give a  method to calculate $b_0$, and for
$b_1$ we refer the interested  readers to  the proof of Theorem~\ref{t11.3.2}.
}
\er

\subsection{Computational procedure of $b_0$}
The procedure to compute the parameter $b_0$ in Assertion (1) of
Theorem~\ref{t11.3.4} is divided into a few steps as follows.

\medskip

{\sc Step 1}. The reduced equations of (\ref{(11.3.28)}) to center
manifold at $\lambda =\Lambda_c$ are expressed by
\begin{equation}
\begin{aligned}
&\frac{dx}{dt}=-\rho y+\frac{1}{<\varphi ,\varphi^*>}<G(x\varphi
+y\psi +\Phi ,\Lambda_c),\varphi^*>,\\
&\frac{dy}{dt}=\rho x+\frac{1}{<\psi ,\psi^*>}<G(x\varphi +y\psi
+\Phi ,\Lambda_c),\psi^*>,
\end{aligned}
\label{(11.3.93)}
\end{equation}
where $\varphi$ and $\psi$ are the eigenvectors of $L_{\lambda}$ at
$\lambda =\Lambda_c, \varphi^*$ and $\psi^*$ the conjugate
eigenvectors, and $L_{\lambda}, G_{\lambda}: H_1\rightarrow H$ the
operators defined by (\ref{(11.3.27)}), $\Phi$ is the center manifold
function.

\medskip

{\sc  Step 2}.  Solving the eigenvectors $\varphi , \psi$ and their
conjugates $\varphi^*, \psi^*$. We know that $\psi_i$ and $\psi^*_i$
are
\begin{equation}
\begin{aligned}
&\varphi =(\xi_1e_K,\xi_2e_K),\ \ \ \ \psi =(\eta_1e_K,
\eta_2e_K),\\
&\varphi^*=(\xi^*_1e_K, \xi^*_2e_K),\ \ \ \ \psi^*=(\eta^*_1e_K,
\eta^*_2e_K),
\end{aligned}
\label{(11.3.94)}
\end{equation}
and $\xi_i, \xi^*_i$ satisfy
\begin{eqnarray*}
&&\left(\begin{array}{cc}
\frac{\Lambda_c\rho_Ku^*_1}{1+\rho_K}-\mu\rho_K-2\alpha
u^{*2}_1&\frac{\alpha
u^*_1}{(1+u^*_3)^2}\\
-\delta u^*_3&-r\rho_K-\delta
u^*_1\end{array}\right)\left(\begin{array}{c}
\xi_1\\
\xi_2\end{array}\right)=\rho\left(\begin{array}{c}
\eta_1\\
\eta_2
\end{array}\right),\\
&&\left(\begin{array}{cc}
\frac{\Lambda_c\rho_Ku^*_1}{1+\rho_K}-\mu\rho_K-2\alpha
u^{*2}_1&\frac{\alpha
u^*_1}{(1+u^*_3)^2}\\
-\delta u^*_3&-r\rho_K-\delta
u^*_1\end{array}\right)\left(\begin{array}{c}
\eta_1\\
\eta_2\end{array}\right)=-\rho\left(\begin{array}{c}
\xi_1\\
\xi_2 \end{array}\right),
\end{eqnarray*}
and \begin{eqnarray*} &&\left(\begin{array}{cc}
\frac{\Lambda_c\rho_Ku^*_1}{1+\rho_K}-\mu\rho_K-2\alpha
u^{*2}_1&-\delta u^*_3\\
\frac{\alpha u^*_1}{(1+u^*_3)^2}&-r\rho_K-\delta
u^*_1\end{array}\right)\left(\begin{array}{c}
\xi^*_1\\
\xi^*_2\end{array}\right)=-\rho\left(\begin{array}{c}
\eta^*_1\\
\eta^*_2\end{array}\right)\\
&&\left(\begin{array}{cc}
\frac{\Lambda_c\rho_Ku^*_1}{1+\rho_K}-\mu\rho_K-2\alpha
u^{*2}_1&-\delta u^*_3\\
\frac{\alpha u^*_1}{(1+u^*_3)^2}&-r\rho_K-\delta
u^*_1\end{array}\right)\left(\begin{array}{l}
\eta^*_1\\
\eta^*_2\end{array}\right)=\rho\left(\begin{array}{c}
\xi^*_1\\
\xi^*_2\end{array}\right),
\end{eqnarray*}
where $\Lambda_c$ is as in (\ref{(11.3.85)}), $e_k$ as in
(\ref{(11.3.37)}), and
\begin{eqnarray*}
&&\frac{\Lambda_c\rho_Ku^*_1}{1+\rho_K}-\mu\rho_K-2\alpha
u^{*2}_1=r\rho_K+\delta
u^*_1,\\
&&\rho =\text{det}\
A^{\Lambda_c}_K=\frac{\alpha\delta_0}{(1+u^*_3)^2}-(\gamma\rho_K+\delta
u^*_1)^2. \end{eqnarray*} Here, we use that
$u^*_1u^*_3=\delta_0/\delta$. From these equations we obtain
\begin{equation}
\begin{array}{ll}
\xi_1=-(r\rho_K+\delta u^*_1),&\xi_2=\delta
u^*_3,\\
\eta_1=-\rho ,&\eta_2=0,\\
\xi^*_1=0,&\xi^*_2=-\rho ,\\
\eta^*_1=\delta u^*_3,&\eta^*_2\gamma\rho_K+\delta u^*_1.
\end{array}\label{(11.3.95)}
\end{equation}
Due to (\ref{(11.3.94)}) and (\ref{(11.3.95)}) we see that
\begin{eqnarray*}
&&<\varphi ,\varphi^*>=<\psi
,\psi^*>=(\eta_1\eta^*_1+\eta_2\eta^*_2)\int_{\Omega}e^2_Kdx=-\delta\rho
u^*_3\int_{\Omega}e^2_Kdx,\\
&&<\varphi ,\psi^*>=<\psi ,\varphi^*>=0.
\end{eqnarray*}

\medskip

{\sc Step 3}.  We need to calculate
$$<G(x\varphi +y\psi +\Phi ,\Lambda_c), \omega^*_j>,\ \ \ \ \text{with}\ \omega^*_1=\varphi^*, \omega_2=\psi^*.$$
By (\ref{(11.3.27)}) we have $G=G_2+G_3$, and
\begin{eqnarray*}
&&G_2(\omega ,\lambda )=\left(\begin{array}{c} -\lambda\nabla
(\omega_1\nabla (-\Delta +I)^{-1}\omega_1)-3\alpha
u^*_1\omega^2_1+\frac{\alpha\omega_1\omega_2}{(1+u^*_3)^2}-\frac{\alpha
u^*_1\omega^2_2}{(1+u^*_3)^3}\\
-\delta\omega_1\omega_2\end{array}\right),\\
&&G_3(\omega ,\lambda )=\left(\begin{array}{c}
-\alpha\omega^3_1-\frac{\alpha\omega_1\omega^2_2}{(1+u^*_3)^3}+\frac{\alpha
u^*_1\omega^3_2}{(1+u^*_3)^4}\\
0\end{array}\right), \end{eqnarray*} for $\omega
=(\omega_1,\omega_2)\in H_1$. By (\ref{(11.3.95)}) we find
$$<G_3(x\varphi +y\psi +\Phi ,\Lambda_c),\varphi^*>=0.$$
Noting that \begin{eqnarray*} &&\int_{\Omega}e_Ke_Je_Idx=0,\ \ \ \
\forall K,J,I\in\mathbb{Z}^2,\\
&&\Phi =(\Phi_1,\Phi_2)=O(x^2), 
\end{eqnarray*}
we have \begin{eqnarray*} &&<G_2(x\varphi +y\psi +\Phi
,\Lambda_c),\varphi^*)\\
&=&\int_{\Omega}[\xi^*_1e_Kg_{21}+\xi^*_2e_Kg_{22}]dx\\
&=&(\text{by}\ \xi^*_1=0)\\
&=&\int_{\Omega}\xi^*_2e_K[-\delta
(x\xi_1e_K+y\eta_1e_K+\Phi_1)(x\xi_2+y\eta_2+\Phi_2)]dx\\
&=&-\delta\xi^*_2\left(\xi_2x\int_{\Omega}\Phi_1e^2_Kdx+\xi_1x\int_{\Omega}\Phi_2e^2_Kdx+\eta_1y\int_{\Omega}\Phi_2e^2_Kdx\right).
\end{eqnarray*}
Thus, we get \begin{eqnarray} &&<G(x\varphi +y\psi +\Phi ,
\Lambda_c),\varphi^*>\label{(11.3.96)}\\
&=&-\delta\xi^*_2\left[\xi_2x\int_{\Omega}\Phi_1e^2_Kdx+\xi_1x\int_{\Omega}\Phi_2e^2_Kdx+\eta_1y\int_{\Omega}\Phi_2e^2_Kdx\right]+o(3).\nonumber
\end{eqnarray}
In the same fashion, we derive 
\begin{align} 
&<G(x\varphi +y\psi
+\Phi , \Lambda_c), \psi^*>\label{(11.3.97)}\\
= &\left(\frac{\alpha\xi_2\eta^*_1}{(1+u^*_3)^2}-6\alpha
u^*_1\xi_1\eta^*_1-\delta\xi_2\eta^*_2\right)x\int_{\Omega}\Phi_1e^2_Kdx
   -6\alpha u^*_1\eta_1\eta^*_1y\int_{\Omega}\Phi_1e^2_Kdx\nonumber\\
&+\left(\frac{\alpha\eta_1\eta^*_1}{(1+u^*_3)^2}-\frac{2\alpha
u^*_1\xi_2\eta^*_1}{(1+u^*_3)^3}-\delta\xi_1\eta^*_2\right)x\int_{\Omega}\Phi_2e^2_Kdx\nonumber\\
&+\left(\frac{\alpha\eta_1\eta^*_1}{(1+u^*_3)^2}-\delta\eta_1\eta^*_3\right)y\int_{\Omega}\Phi_2e^2_Kdx
+\frac{\Lambda_c\xi_1\eta^*_1}{1+\rho_K}x\int_{\Omega}\Phi_1|\nabla
e_K|^2dx\nonumber\\
&+\frac{\Lambda_c\eta_1\eta^*_1}{1+\rho_K}y\int_{\Omega}\Phi_1|\nabla
e_K|^2dx -\frac{1}{2}\Lambda_c\xi_1\eta^*_1x\int_{\Omega}e^2_K\Delta
(-\Delta +I)^{-1}\Phi_1dx\nonumber\\
&-\frac{1}{2}\Lambda_c\eta_1\eta^*_1y\int_{\Omega}e^2_K\Delta
(-\Delta +I)^{-1}\Phi_1dx\nonumber\\
&-\left[\eta^*_1\left(\frac{\alpha
u^*_1}{(1+u^*_3)^4}\xi^3_2-\frac{\alpha\xi^2_2\xi_1}{(1+u^*_3)^3}-\alpha\xi^3_1\right)\int_{\Omega}e^4_Kdx\right]x^3\nonumber\\
&-\left[\eta^*_1\left(\frac{\alpha\xi^2_2\eta_1}{(1+u^*_3)^3}+3\alpha\xi^2_1\eta_1\right)\int_{\Omega}e^4_Kdx\right]x^2y\nonumber\\
&-\left(3\alpha\xi_1\eta^2_1\eta^*_1\int_{\Omega}e^4_Kdx\right)xy^2
-\left(\alpha\eta^3_1\eta^*_1\int_{\Omega}e^4_Kdx\right)y^3+o(3).\nonumber
\end{align}

{\sc Step 4}. By the formula of center manifold function in the complex case in  
Theorem A.1 in \cite{MW09c}, we have 
\begin{equation} \Phi
=\left(\begin{array}{c} \Phi_1\\
\Phi_2\end{array}\right)=\left(\begin{array}{c} \Phi^1_1\\
\Phi^1_2\end{array}\right)+\left(\begin{array}{c} \Phi^2_1\\
\Phi^2_2\end{array}\right)+\left(\begin{array}{c} \Phi^3_1\\
\Phi^3_2\end{array}\right)+o(3),\label{(11.3.98)} 
\end{equation} 
with
\begin{equation}
\begin{aligned}
&-L_{\lambda_c}\left(\begin{array}{c} \Phi^1_1\\
\Phi^1_2\end{array}\right)=x^2G_{11}+xy(G_{12}+G_{21})+y^2G_{22},\\
&-(L^2_{\lambda_c}+4\rho^2)L_{\lambda_c}\left(\begin{array}{c}
\Phi^2_1\\
\Phi^2_2\end{array}\right)=2\rho^2\left[(x^2-y^2)(G_{22}-G_{11})-2xy(G_{12}+G_{21})\right],\\
&(L^2_{\lambda_c}+4\rho^2)\left(\begin{array}{c} \Phi^3_1\\
\Phi^3_2\end{array}\right)=\rho\left[(y^2-x^2)(G_{12}+G_{21})+2xy(G_{11}-G_{22})\right].
\end{aligned}
\label{(11.3.99)}
\end{equation}
Here $G_{ij}=G_2(\Psi^i,\Psi^j,\lambda_c)$ with $\Psi^1=\varphi$
and $\Psi^2=\psi$, and $G_2$ is as defined in Step 3. Namely
$$
G_{ij}=\left(\begin{array}{c} -\Lambda_c\nabla (\Psi^i_1\nabla
(-\Delta +I)^{-1}\Psi^j_1)\\
0\end{array}\right)
+\left(\begin{array}{c}
\frac{\alpha\Psi^i_1\Psi^j_2}{(1+u^*_3)^2}-\frac{\alpha
u^*_1\Psi^i_2\Psi^j_2}{(1+u^*_3)^2}-3\alpha u^*_1\Psi^i_1\Psi^j_1\\
-\delta\Psi^i_1\Psi^j_2\end{array}\right),
$$
with $\Psi^i_l=\Gamma^i_le_K, 1\leq i, l\leq 2$, and
\begin{equation}
\Gamma^1_1=\xi_1,\ \ \ \ \Gamma^1_2=\xi_2,\ \ \ \
\Gamma^2_1=\eta_1,\ \ \ \ \Gamma^2_2=\eta_2,\label{(11.3.100)}
\end{equation}
which are given by (\ref{(11.3.95)}).

Direct calculation shows that 
\begin{eqnarray}
G_{ij}&=&-\frac{\Lambda_c\Gamma^i_1\Gamma^j_1}{1+\rho_K}\nabla
(e_K\nabla e_K)\left(\begin{array}{c} 1\\
0\end{array}\right)\label{(11.3.101)}\\
&&+e^2_K\left(\begin{array}{c}
\frac{\alpha\Gamma^i_1\Gamma^j_2}{(1+u^*_3)^2}-\frac{\alpha
u^*_1\Gamma^i_2\Gamma^j_2}{(1+u^*_3)^3}-3\alpha
u^*_1\Gamma^i_1\Gamma^j_1\\
-\delta\Gamma^i_1\Gamma^j_2\end{array}\right).\nonumber
\end{eqnarray}
For simplicity, we only consider the case where $K=(K_1,0)$. In this
case, by (\ref{(11.3.37)}) we can see that
$$
e^2_K=\frac{1}{2}(e_0+e_{2K}),\ \ \ \ \nabla (e_K\nabla
e_K)=-\rho_Ke_{2K}.
$$ 
Then, by (\ref{(11.3.101)}),  we have
\begin{equation}
G_{ij}=\left(\begin{array}{c} h^0_{ij}\\
g^0_{ij}\end{array}\right)e_0+\left(\begin{array}{c} h^{2K}_{ij}\\
g^{2K}_{ij}\end{array}\right)e_{2K},\ \ \ \ 1\leq i, j\leq
2,\label{(11.3.102)}
\end{equation}
where
\begin{equation}
\begin{aligned}
&h^0_{ij}=\frac{1}{2}\left[\frac{\alpha\Gamma^i_1\Gamma^j_2}{(1+u^*_3)^2}-\frac{\alpha
u^*_1\Gamma^i_2\Gamma^j_2}{(1+u^*_3)^3}-3\alpha
u^*_1\Gamma^i_1\Gamma^j_1\right],\\
&h^{2K}_{ij}=\frac{\rho_K\lambda_c\Gamma^i_1\Gamma^j_1}{1+\rho_K}+h^0_{ij},\\
&g^0_{ij}=g^{2K}_{ij}=-\frac{1}{2}\delta\Gamma^i_1\Gamma^j_2.
\end{aligned}
\label{(11.3.103)}
\end{equation}
Let
\begin{equation}
\left(\begin{array}{c} \Phi^K_1\\
\Phi^K_2\end{array}\right)=\left(\begin{array}{c} \varphi^0_{k1}\\
\varphi^0_{k2}\end{array}\right)e_0+\left(\begin{array}{c}
\varphi^{2K}_{k1}\\
\varphi^{2K}_{k2}\end{array}\right)e_{2K},\ \ \ \ 1\leq k\leq
3.\label{(11.3.104)}
\end{equation}
Then it follows from (\ref{(11.3.99)}) and (\ref{(11.3.102)}) that
\begin{equation}
\begin{aligned}
&\left(\begin{array}{c} \varphi^0_{11}\\
\varphi^0_{12}\end{array}\right)=B^{-1}_0\left[x^2\left(\begin{array}{c}
h^0_{11}\\
g^0_{11}\end{array}\right)+xy\left(\begin{array}{c}
h^0_{12}+h^0_{21}\\
g^0_{12}+g^0_{21}\end{array}\right)+y^2\left(\begin{array}{c}
h^0_{22}\\
g^0_{22}\end{array}\right)\right],\\
&\left(\begin{array}{c} \varphi^{2K}_{11}\\
\varphi^{2K}_{12}\end{array}\right)=B^{-1}_{2K}\left[x\left(\begin{array}{c}
h^{2K}_{11}\\
g^{2K}_{11}\end{array}\right)+xy\left(\begin{array}{c}
h^{2K}_{12}+h^{2K}_{21}\\
g^{2K}_{12}+g^{2K}_{21}\end{array}\right)+y^2\left(\begin{array}{c}
h^{2K}_{22}\\
g^{2K}_{22}\end{array}\right)\right],\\
&\left(\begin{array}{c} \varphi^0_{21}\\
\varphi^0_{22}\end{array}\right)=2\rho^2B^{-1}_0(B^2_0+4\rho^2I)^{-1}\left[(x^2-y^2)\left(\begin{array}{c}
h^0_{22}-h^0_{11}\\
g^0_{22}-g^0_{11}\end{array}\right)-2xy\left(\begin{array}{c}
h^0_{12}+h^0_{21}\\
g^0_{12}+g^0_{21}\end{array}\right)\right],\\
&\left(\begin{array}{c} \varphi^{2K}_{21}\\
\varphi^{2K}_{22}\end{array}\right)=2\rho^2B^{-1}_{2K}(B^2_{2K}+4\rho^2I)^{-1}\left[(x^2-y^2)\left(\begin{array}{c}
h^{2K}_{22}-h^{2K}_{11}\\
g^{2K}_{22}-g^{2K}_{11}\end{array}\right)-2xy\left(\begin{array}{c}
h^{2K}_{12}+h^{2K}_{21}\\
g^{2K}_{12}+g^{2K}_{21}\end{array}\right)\right],\\
&\left(\begin{array}{c} \varphi^0_{31}\\
\varphi^0_{32}\end{array}\right)=\rho
(B^2_0+4\rho^2I)^{-1}\left[(y^2-x^2)\left(\begin{array}{c}
h^0_{12}+h^0_{21}\\
g^0_{12}+g^0_{21}\end{array}\right)+2xy\left(\begin{array}{c}
h^0_{11}-h^0_{22}\\
g^0_{11}-g^0_{22}\end{array}\right)\right],\\
&\left(\begin{array}{c} \varphi^{2K}_{31}\\
\varphi^{2K}_{32}\end{array}\right)=\rho
(B^2_{2K}+4\rho^2I)^{-1}\left[(y^2-x^2)\left(\begin{array}{c}
h^{2K}_{12}+h^{2K}_{21}\\
g^{2K}_{12}+g^{2K}_{21}\end{array}\right)+2xy\left(\begin{array}{c}
h^{2K}_{11}-h^{2K}_{22}\\
g^{2K}_{11}-g^{2K}_{22}\end{array}\right)\right],
\end{aligned}
\label{(11.3.105)}
\end{equation}
where $B_k=-A^{\lambda_c}_k$ with $A^{\lambda_c}_k$ as defined by
(\ref{(11.3.83)}). By (\ref{(11.3.98)}) and (\ref{(11.3.104)}) we obtain an 
explicit expression of $\Phi$ as follows:
\begin{equation}
\begin{aligned}
&\Phi_1=(\varphi^0_{11}+\varphi^0_{21}+\varphi^0_{31})e_0+(\varphi^{2K}_{11}+\varphi^{2K}_{21}+\varphi^{2K}_{31})e_{2K}+o(2),\\
&\Phi_2=(\varphi^0_{12}+\varphi^0_{22}+\varphi^0_{32})e_0+(\varphi^{2K}_{12}+\varphi^{2K}_{22}+\varphi^{2K}_{32})e_{2K}+o(2).
\end{aligned}
\label{(11.3.106)}
\end{equation}
Here, by (\ref{(11.3.105)}), (\ref{(11.3.103)}), and
(\ref{(11.3.100)}), $\varphi^k_{ij}$ are 2-order homogeneous
functions of $(x,y)$, with the coefficients depending  explicitly
on the parameters  defined in (\ref{(11.3.20)}).

\medskip

{\sc  Step 5}.  Finally, inserting (\ref{(11.3.106)}) into
(\ref{(11.3.96)}) and (\ref{(11.3.97)}), we can write (\ref{(11.3.93)})
in the following form
\begin{eqnarray*}
&&\frac{dx}{dt}=-\rho
y+a_{11}x^3+a_{12}x^2y+a_{13}xy^2+a_{14}y^3+o(4),\\
&&\frac{dy}{dt}=\rho
x+a_{21}x^3+a_{22}x^2y+a_{23}xy^2+a_{24}y^3+o(4).
\end{eqnarray*}
Then, by Theorem~2.4.5 in \cite{ptd}, 
the parameter $b_0$ in Theorem~\ref{t11.3.4} is
obtained by
$$b_0=3a_{11}+3a_{24}+a_{12}+a_{23},$$
where $a_{11},a_{24},a_{12},a_{23}$ can be explicitly expressed in
the terms in (\ref{(11.3.95)})-(\ref{(11.3.97)}).

\subsection{Transition for the system (\ref{(11.3.21)})}

We are now in a position to discuss the transition of
(\ref{(11.3.21)}). With the translation (\ref{(11.3.24)}), the system
(\ref{(11.3.21)}) is rewritten in the following form
\begin{equation}
\begin{aligned}
&\frac{\partial u_1}{\partial t}=\mu\Delta u_1-2\alpha
u^{*2}_1u_1-u^*_1\Delta u_2+\frac{\alpha
u^*_1u+3}{(1+u^*_3)^2}+g(u),\\
&\frac{\partial u_2}{\partial t}=\Delta u_2-u_2+\lambda u_1,\\
&\frac{\partial u_3}{\partial t}=r\Delta u_3-\delta u^*_1u_3-\delta
u^*_3u_1-\delta u_1u_3,\\
&\left.\frac{\partial u}{\partial n}\right|_{\partial\Omega}=0,\\
&u(0)=u_0,
\end{aligned}
\label{(11.3.107)}
\end{equation}
where $g(u)$ is as in (\ref{(11.3.26)}). Here the notation $u$ stands for three-component unknown:
$$u=(u_1, u_2, u_3).$$

Let
$$
L_{\lambda}u=\left(\begin{array}{ccc}
\mu\Delta -2\alpha u^{*2}_1&-u^*_1\Delta&\frac{\alpha
u^*_1}{(1+u^*_3)^2}\\
\lambda&\Delta -1&0\\
-\delta u^*_3&0&r\Delta -\delta
u^*_1\end{array}\right)\left(\begin{array}{c} u_1\\
u_2\\
u_3\end{array}\right).
$$ 
Then, all eigenvalues $\beta^j_k(\lambda )$
and eigenvectors $\psi^j_k$ of $L_{\lambda}$ satisfy
$$D^{\lambda}_k\left(\begin{array}{c}
\xi^j_{k1}\\
\xi^j_{k2}\\
\xi^j_{k3}\end{array}\right)=\beta^j_k(\lambda
)\left(\begin{array}{c} \xi^j_{k1}\\
\xi^j_{k2}\\
\xi^j_{k3}\end{array}\right),\ \ \ \ 1\leq j\leq 3,\
k\in\mathbb{Z}^2,$$ with
$$\psi^j_k=(\xi^j_{k1}e_k,\xi^j_{k2}e_k,\xi^j_{k3}e_k),$$
and $e_k$ as in (\ref{(11.3.37)}), $D^{\lambda}_k$ is a $3\times 3$ matrix
given by
$$D^{\lambda}_k=\left(\begin{array}{ccc}
-(\mu\rho_k+2\alpha u^{*2}_1)&u^*_1\rho_k&\frac{\alpha
u^*_1}{(1+u^*_3)^2}\\
\lambda&-(\rho_k+1)&0\\
-\delta u^*_3&0&-(r\rho_k+\delta u^*_1)\end{array}\right).$$

We introduce the following three parameters:
\begin{eqnarray*}
&&A^{\lambda}_k=-\text{tr}D^{\lambda}_k=\mu\rho_k+2\alpha
u^{*2}_1+\rho_k+1+r\rho_k+\delta u^*_1,\\
&&B^{\lambda}_k=\text{det}\left(\begin{array}{cc}
-(\mu\rho_k+2\alpha u^{*2}_1)&u^*_1\rho_k\\
\lambda&-(\rho_k+1)\end{array}\right)+\text{det}\left(\begin{array}{cc}
-(\mu\rho_k+2\alpha u^{*2}_1)&\frac{\alpha u^*_1}{(1+u^*_3)^2}\\
-\delta u^*_3&-(r\rho_k+\delta u^*_1)\end{array}\right)\\
&&\ \ \ \ \ \ \ \ +(\rho_k+1)(r\rho_k+\delta u^*_1), \\
&&C^{\lambda}_k=-\text{det}D^{\lambda}_k=(\mu\rho_k+2\alpha
u^{*2}_1)(\rho_k+1)(r\rho_k+\delta u^*_1)\\
&&~~~~~~~~~~~\ \ \ \ \ \ \ \ \ \ \ -u^*_1\rho_k\lambda
(r\rho_k+\delta u^*_1)+\frac{\alpha u^*_1}{(1+u^*_3)^2}\delta
u^*_3(\rho_k+1).
\end{eqnarray*}
By the Routh-Hurwitz theorem, we know that
all eigenvalues $\beta^j_k$ of $D^j_k$ have negative real parts if
and only if
\begin{equation}
A^{\lambda}_k>0,\ \ \ \ A^{\lambda}_kB^{\lambda}_k-C^{\lambda}_k>0,\
\ \ \ C^{\lambda}_k>0.\label{(11.3.108)}
\end{equation}

Let $\Lambda^c$ and $K=(K_1,K_2)$ satisfy
\begin{equation}
\begin{aligned}
&A^{\Lambda_c}_K>0,\ \ \ \
A^{\Lambda_c}_KB^{\Lambda_c}_K-C^{\lambda_c}_K=0,\ \ \ \
C^{\Lambda_c}_K>0,\\
&A^{\Lambda_c}_k>0,\ \ \ \
A^{\Lambda_c}_kB^{\Lambda_c}_k-C^{\Lambda_c}_k>0,\ \ \ \
C^{\Lambda_c}_k>0,\ \ \ \ \forall k\ \text{with}\
\rho_k\neq\rho_K.
\end{aligned}
\label{(11.3.109)}
\end{equation}
Then $\Lambda_c$ satisfies that
\begin{align}
\Lambda_c=&\inf_{\rho_k}\frac{1}{\rho_ku^*_1}[(\mu
+r)\rho_k+2\alpha u^{*2}_1+\delta u^*_1]\label{(11.3.110)}\\
&\times\left[(r+1)\rho_k+\delta
u^*_1+1+\frac{\alpha\delta}{(\mu\rho_k+2\alpha
u^{*2}_1+\rho_k+1)(1+u^*_3)^2}\right],\nonumber
\end{align}
and $\rho_K$ satisfies (\ref{(11.3.110)}). In particular, under
the condition (\ref{(11.3.109)}), there is a pair of complex
eigenvalues $\beta^1_K(\lambda )$ and $\beta^2_K(\lambda )$ of
$D^{\lambda}_K,$ such that
\begin{equation}
\text{Re}\ \beta^{1,2}_K(\lambda )\left\{\begin{array}{ll} <0,&\lambda
<\Lambda_c,\\
=0,&\lambda =\Lambda_c,\\
>0,&\lambda >\Lambda_c,
\end{array}\right.\label{(11.3.111)}
\end{equation}
and the other eigenvalues $\beta^j_k(\lambda )$ of $L_{\lambda}$
satisfy
\begin{equation}
\left\{\begin{array}{l} \text{Re}\ \beta^j_k(\Lambda_c)<0,\ \ \ \
\forall k\ \text{with}\ \rho_k\neq\rho_K,\ \text{and}\ 1\leq j\leq
3,\\
\beta^3_K(\Lambda_c)<0.
\end{array}
\right.\label{(11.3.112)}
\end{equation}

Let $\lambda_c$ and $K^*=(K^*_1,K^*_2)$ satisfy
\begin{equation}
\begin{aligned}
&A^{\lambda_c}_{K^*}>0,\ \ \ \
A^{\lambda_c}_{K^*}B^{\lambda_c}_{K^*}-C^{\lambda_c}_{K^*}>0,\ \ \ \
C^{\lambda_c}_{K^*}=0,\\
&A^{\lambda_c}_k>0,\ \ \ \
A^{\lambda_c}_kB^{\lambda_c}_k-C^{\lambda_c}_k>0,\ \ \ \
C^{\lambda_c}_k>0,\ \ \ \ \forall k\ \text{with}\
\rho_k\neq\rho_{K^*}.\end{aligned} \label{(11.3.113)}
\end{equation}
Then $\lambda_c$ is given by
\begin{equation}
\lambda_c=\inf_{\rho_k}\frac{(\rho_k+1)}{\rho_k u^*_1}\left[\mu\rho_k+2\alpha
u^{*2}_1+\frac{\alpha\delta}{(1+u^*_3)^2(r\rho_k+\delta
u^*_1)}\right],\label{(11.3.114)}
\end{equation}
and $\lambda_c$ arrives its minimal at $\rho_{K^*}$. From the
Routh-Hurwitz criterion (\ref{(11.3.108)}),  we deduce that with
(\ref{(11.3.113)}) there is a real eigenvalue $\beta^1_{K^*}(\lambda
)$ of $D^{\lambda_c}_{K^*}$ satisfies
\begin{align}
&
\beta_{K^*}(\lambda )\left\{\begin{array}{ll} <0,&\lambda
<\lambda_c,\\
=0,&\lambda =\lambda_c,\\
>0,&\lambda >\lambda_c,
\end{array}\right.\label{(11.3.115)}
\\
& 
\left\{\begin{array}{l} \text{Re}\beta^j_{K^*}(\lambda_c)<0,\ \ \ \
j=2,3,\\
\text{Re}\beta^j_k(\lambda_c)<0,\ \ \ \ \forall k\in\mathbb{Z}^2\
\text{with}\ \rho_k\neq\rho_{K^*}\ \text{and}\ 1\leq j\leq 3.
\end{array}\right.\label{(11.3.116)}
\end{align}

It is clear that (\ref{(11.3.111)}) and (\ref{(11.3.112)}) hold true
as $\Lambda_c<\lambda_c$, and (\ref{(11.3.114)})-(\ref{(11.3.115)})
hold true as $\lambda_c<\Lambda_c$. Hence, we have the following
transition theorem for (\ref{(11.3.107)}).

\bt\la{t11.3.5}
 Let $\Lambda_c$ and $\lambda_c$ be given by
(\ref{(11.3.110)}) and (\ref{(11.3.114)}) respectively. Then,
Assertions (1) and (2) of Theorem~\ref{t11.3.4} hold true for the system
(\ref{(11.3.107)}).
\et

\section{Biological Conclusions}
\subsection{Biological significance of transition theorems}
Pattern formation is one of the characteristics for bacteria
chemotaxis, and is fully characterized by the  dynamic transitions. 
Theorems~\ref{t11.3.1}--\ref{t11.3.5} tell us that  
the nondimensional
parameter $\lambda$, given by
\begin{equation}
\lambda =\frac{\sqrt{\alpha_2}r_1\chi}{r_2k_2},\label{(11.3.119)}
\end{equation}
plays a crucial role to determine the dynamic transition and pattern formation. 
Actually, the key factor in (\ref{(11.3.119)}) is the product
of the chemotactic coefficient $\chi$ and the production rate
$r_1:\chi r_1$, which depends on the type of bacteria. When
$\lambda$ is less than some critical value $\lambda_c$, the uniform
distribution of biological individuals is a stable state. When $\lambda$ exceeds $\lambda_c$, the bacteria cells aggregate to form
more complex  and stable patterns.

As seen in (\ref{(11.3.43)}), (\ref{(11.3.85)}), (\ref{(11.3.86)}) and
(\ref{(11.3.110)}), under different biological conditions, the critical
parameter $\lambda_c$ takes  different forms and values. But, a
general formula for $\lambda_c$ is of the following type:
\begin{equation}
\lambda_c=a_0+\inf_{\rho_k}\left(a_1\rho_k+\frac{a_2}{\rho_k}+\frac{a_3}{b_1\rho_k+b_0}+\frac{a_4}{\rho_k(b_1\rho_k+b_0)}\right),\label{(11.3.120)}
\end{equation}
where $\rho_k$ are taken as the eigenvalues of $-\Delta$ with the
Neumann boundary condition. When $\Omega$ is a rectangular region,
$\rho_k$ are given by (\ref{(11.3.37)}), and the coefficients
$a_j$  $(1\leq j\leq 4), b_0,b_1\geq 0$ depend on the parameters in 
(\ref{(11.3.20)}), with
$$a_0,a_1,a_2,b_0,b_1>0,\ \ \ \ a_3,a_4\geq 0.$$
In particular, for the system  with rich nutrient supplies,
(\ref{(11.3.120)}) becomes
$$\lambda_c=a_0+\inf_{\rho_k}\left[a_1\rho_k+\frac{a_2}{\rho_k}\right].$$

The eigenvalues $\rho_k$, depending on the geometry of $\Omega$,
satisfy
\begin{equation}
\left\{\begin{array}{l}
0=\rho_0<\rho_1\leq\cdots\leq\rho_k\leq\cdots ,\ \ \ \
\rho_k\rightarrow\infty\ \text{as}\ k\rightarrow\infty ,\\
\rho_1\propto\frac{1}{L^2},
\end{array}\right.\label{(11.3.121)}
\end{equation}
where $L$ is the length  scale  of $\Omega$.

We infer from (\ref{(11.3.120)}) and (\ref{(11.3.121)}) that
$$
\lambda_c\rightarrow\infty\ \ \ \ \text{as}\ \ \ \ |\Omega
|\rightarrow 0\ \ \ \ (L\rightarrow 0).
$$ 
It implies that when the container $\Omega$ is small, the homogenous state is state and there is 
no pattern formation of bacteria under any biological
conditions.

\subsection{Spatiotemporal  oscillation}
Theorems~\ref{t11.3.4} and \ref{t11.3.5} show that there are two critical parameters
$\lambda_c$ and $\Lambda_c$, such that if $\lambda_c<\Lambda_c$,  the
patterns formed by biological organisms are steady, as
exhibited by many  experimental results, and if
$\Lambda_c<\lambda_c$ a spatial-temporal oscillatory behavior  takes place.

For the case with rich nutrient,
$$u^*_1=1,\ \ \ \ u^*_3=\infty .$$
In this situation, $\lambda_c$ in (\ref{(11.3.86)}) is reduced to
(\ref{(11.3.40)}), and obviously we have that
$$\lambda_c<\Lambda_c\ \ \ \ \text{for\ both\ (\ref{(11.3.85)})\ and\
(\ref{(11.3.110)}),}
$$
and the dynamic transition and pattern formation are determined by Theorems~\ref{t11.3.1}
and \ref{t11.3.2}. Hence there is no spatiotemporal oscillations  for the rich nutrient case, and 
 the time periodic oscillation of chemotaxis 
occurs only for the case where the nutrient is
moderately  supplied.

In particular, if $\mu ,r\cong 0$, and
$$\delta^2u^{*2}_1(1+u^*_3)^2<\alpha\delta_0,$$
then for $\Lambda_c$ defined by (\ref{(11.3.85)}) and
(\ref{(11.3.110)}), we have
$$\Lambda_c<\lambda_c.$$
In this case, a spatial-temporal oscillation pattern are expected for 
$\lambda >\Lambda_c$.

\subsection{Transition types} One of the  most important aspects of the study for phase transitions is 
to determine  the transition types for a given system.  The main theorems in this article provide precise 
information  on the transition types. In all cases, types are precisely determined by the sign of 
some non dimensional parameters; see $b$, $b_0$  and$b_1$ respectively in the main theorems. 
Hence a global phase diagram can be obtained easily by setting the related parameter to be zero. 

For example, when $\Omega =(0,L_1)$ is one-dimensional  or when $K=(K_1,0)$
(resp. $K=(0,K_2))$, the parameter $b$ in (\ref{(11.3.44)}) can be
simplified into the following form
\begin{equation}
b=2\left[-3\mu\rho_K+9\alpha -\frac{(2\mu\rho_K+\alpha
)(2\mu\lambda^2_K+28\alpha\rho_K+4\alpha
-\mu\rho_K)}{(\mu\rho_{2K}+2\alpha
)(\rho_{2K}+1)-\rho_{2K}\lambda_c}\right].\label{(11.3.122)}
\end{equation}

For a non-growth system, $\alpha =0, K=(1,0), \lambda_c=\mu
(\rho_K+1)$. Then, (\ref{(11.3.122)}) becomes
\begin{equation}
b=\frac{\mu}{3}(1-20\lambda_1),\ \ \ \
\lambda_1=\frac{\pi^2}{L^2_1},\label{(11.3.123)}
\end{equation}
and $\lambda =\frac{ar_1x}{r_2k_2}$, with $a=\frac{1}{|\Omega
|}\int_{\Omega}u_1dx$. It follows from (\ref{(11.3.123)}) that
\begin{equation}
b\left\{\begin{aligned}
&  <0 &&\text{ if }\ L_1<2\sqrt{5}\pi ,\\
& >0 &&\text{ if }\ L_1>2\sqrt{5}\pi .
\end{aligned}\right.\label{(11.3.124)}
\end{equation}
By Theorems~\ref{t11.3.1} and \ref{t11.3.2}, the phase transition of (\ref{(11.3.33)})
and (\ref{(11.3.29)}) from $(u,\lambda )=(u^*,\lambda_c)$ is
continuous if the length scale $L_1$ of $\Omega$ is less than
$2\sqrt{5}\pi$, and jump if   $L_1$ is bigger than $2\sqrt{5}\pi$. 

In addition, when we take 
$$\chi (u)=\frac{\chi_1u_1}{(\beta +u_2)^2}$$ 
as   the chemotaxis function, by Remark~\ref{r11.3.5}, the parameter $b$ of (\ref{(11.3.123)}) is replaced by
$$
b_1=\frac{\mu}{3}\left(1-\frac{20\pi^2}{L^2_1}\right)+\frac{4\kappa\mu^2\pi^2}{L^2_1}
$$
with
$$\kappa =\frac{k_2}{\beta\chi +k_2\lambda_c},\ \ \ \ \text{and}\
\lambda_c=\mu\left(\frac{\pi^2}{L^2_1}+1\right).$$

The above conclusion amounts to saying that for a non-growth system,
the parameter
$$\lambda =\frac{r_1\chi}{r_2k_2}a,\ \ \ \ \text{with}\
a=\frac{1}{|\Omega |}\int_{\Omega}u_1dx,$$ is proportional to the
average density a of initial condition of $u_1(u_1$ is
conservation). Hence, the biological individual is in a homogenous
distribution state provided
$$
\frac{1}{|\Omega |}\int_{\Omega}\varphi
dx<\frac{r_2k_2}{r_1\chi}\mu\left(\frac{\pi^2}{L^2_1}+1\right),\ \ \
\ \varphi =u_1(0),
$$ 
and the bacteria will aggregate to form numbers
of high density regions provided
\begin{equation}
\frac{1}{|\Omega |}\int_{\Omega}\varphi
dx>\frac{r_2k_2}{r_1\chi}\mu\left(\frac{\pi^2}{L^2_1}+1\right).\label{(11.3.125)}
\end{equation}
Moreover, under the condition (\ref{(11.3.125)}), if the scale $L_1$
of $\Omega$ is smaller than some critical value $L_c$ (in
(\ref{(11.3.124)}) $L_c=2\sqrt{5}\pi$), i.e. $L_1<L_c$, the
continuous transition implies that there is only one high density
region of bacteria to be formed, and if $L_1>L_c$ then the jump
transition expects a large number of high density regions to appear.

\subsection{Pattern formation}
As mentioned before, the pattern formation behavior is dictated by the dynamic transition of the system. In this article, we studied the formation of two type patterns--the lamella and the rectangular patterns, although the approach can be generalized to study  the formation of other more complex patterns. 

For a growth system, the critical parameter $\lambda_c$ takes its
value at some eigenvalue $\rho_K$ of $-\Delta$ for $K=(K_1,K_2)$,
as shown by (\ref{(11.3.43)}) and (\ref{(11.3.86)}). 
From the pattern formation point of view, for the Type-I transition, the patterns described by the transition solutions in thee main theorems are either lamella or rectangular:
\begin{align*}
& \text{lamella pattern} && \text{  for } K_1K_2=0, \\
& \text{rectangular pattern } && \text { for } K_1 K_2 \not=0.
\end{align*}
In the case where $b>0$,  the system undergoes a more drastic change. 
As $\lambda^\ast < \lambda < \lambda_c$,  the homogeneous state, 
the new patterns
$v_2^\lambda$ and $v_4^\lambda$ are metastable. 
For $\lambda> \lambda_c$, the system undergoes transitions to more complex patterns away from the basic homogeneous state.
\bibliographystyle{siam}

\begin{thebibliography}{10}

\bibitem{brenner}
{\sc M.~P. Brenner, L.~S. Levitov, and E.~O. Budrene}, {\em Physical mechanisms
  for chemotactic pattern formation by bacteria}, Biophysical Journal, 74
  (1998), pp.~1677--1693.

\bibitem{BB91}
{\sc E.~O. Budrene and H.~C. Berg}, {\em Complex patterns formed by motile
  cells of {E}scherichia coli}, Nature, 349 (1991), pp.~630--633.

\bibitem{BB95}
\leavevmode\vrule height 2pt depth -1.6pt width 23pt, {\em Dynamics of
  formation of symmetric patterns of chemotactic bacteria}, Nature, 376 (1995),
  pp.~49--53.

\bibitem{guo10}
{\sc Y.~Guo and H.~J. Hwang}, {\em Pattern formation ({I}): the
  {K}eller-{S}egel model}, J. Differential Equations, 249 (2010),
  pp.~1519--1530.

\bibitem{KS70}
{\sc E.~F. Keller and L.~A. Segel}, {\em Initiation of slime mold aggregation
  viewed as an instability}, J. Theor. Biol., 26 (1970), pp.~399--415.

\bibitem{LSW11}
{\sc H.~Liu, T.~Sengul, and S.~Wang}, {\em Dynamic transitions for quasilinear
  systems and cahn-hilliard equation with onsager mobility}, Journal of
  Mathematical Physics,  (2011).

\bibitem{ptd}
{\sc T.~Ma and S.~Wang}, {\em Phase Transition Dynamics in Nonlinear Sciences},
  submitted, 2009.

\bibitem{MW09c}
\leavevmode\vrule height 2pt depth -1.6pt width 23pt, {\em Dynamic transition
  theory for thermohaline circulation}, Physica D, 239:3-4 (2010),
  pp.~167--189.

\bibitem{ruan}
{\sc P.~Magal and S.~Ruan}, {\em Center manifolds for semilinear equations with
  non-dense domain and applications to {H}opf bifurcation in age structured
  models}, Mem. Amer. Math. Soc., 202 (2009), pp.~vi+71.

\bibitem{murray}
{\sc J.~Murray}, {\em Mathematical Biology, II}, 3rd Ed. Springer-Verlag, 2002.

\bibitem{perthame08}
{\sc G.~Nadin, B.~Perthame, and L.~Ryzhik}, {\em Traveling waves for the
  {K}eller-{S}egel system with {F}isher birth terms}, Interfaces Free Bound.,
  10 (2008), pp.~517--538.

\bibitem{perthame09}
{\sc B.~Perthame and A.-L. Dalibard}, {\em Existence of solutions of the
  hyperbolic {K}eller-{S}egel model}, Trans. Amer. Math. Soc., 361 (2009),
  pp.~2319--2335.

\bibitem{perthame11}
{\sc B.~Perthame, C.~Schmeiser, M.~Tang, and N.~Vauchelet}, {\em Travelling
  plateaus for a hyperbolic {K}eller-{S}egel system with attraction and
  repulsion: existence and branching instabilities}, Nonlinearity, 24 (2011),
  pp.~1253--1270.

\end{thebibliography}

\end{document}